\documentclass[pra,twocolumn,superscriptaddress,notitlepage,showpacs,showkeys]{revtex4-1}
\usepackage{graphicx,amsmath,amsfonts,amssymb,upgreek,txfonts,color,physics}
\usepackage{hyperref}
\hypersetup{colorlinks,linkcolor=blue,citecolor=red,urlcolor=blue,breaklinks=true}
\usepackage{tabularx}
\usepackage[utf8x]{inputenc} 
\usepackage{color}
\usepackage{import}
\usepackage{enumitem}
\usepackage{lipsum}
\usepackage{mathtools}
\usepackage[dvipsnames]{xcolor}
\usepackage{newunicodechar}
\usepackage{natbib}

\begin{document}

\title{Improving photon blockade, entanglement and mechanical-cat-state generation in a generalized cross-Kerr optomechanical circuit}

\author{H. Solki} 
\address{Department of Physics, University of Isfahan, Hezar-Jerib, Isfahan 81746-73441, Iran}
\address{Center of Quantum Science and Technology (CQST), University of Isfahan, Hezar-Jerib, Isfahan 81746-73441, Iran}

\author{Ali Motazedifard}
\email{motazedifard.ali@gmail.com}
\address{Quantum Sensing Lab, Quantum Metrology Group, Iranian Center for Quantum Technologies (ICQT), Tehran, Tehran 15998-14713, Iran}
\address{Quantum Optics Group, Department of Physics, University of Isfahan, Hezar-Jerib, Isfahan 81746-73441, Iran}



\author{M. H. Naderi}
\email{mhnaderi@phys.ui.ac.ir}
\address{Department of Physics, University of Isfahan, Hezar-Jerib, Isfahan 81746-73441, Iran}
\address{Quantum Optics Group, Department of Physics, University of Isfahan, Hezar-Jerib, Isfahan 81746-73441, Iran}

\date{\today}

\begin{abstract}
We propose a feasible experimental scheme to improve the few-photon optomechanical effects, including photon blockade and mechanical-Schr{\"o}dinger  cat-state generation, as well as photon-phonon entanglement in a tripartite microwave optomechanical circuit. The system under consideration is formed by a single-Cooper-pair transistor, a microwave \textit{LC} resonator, and a micromechanical resonator. Our scheme is based on an additional higher-order (generalized) nonlinear cross-Kerr type of coupling, linearly dependent on photon number while quadratically dependent on mechanical phonon one, which can be realized via adjusting the gate charge of the Cooper-pair transistor. We show, both analytically and numerically, that the presence of both cross-Kerr and generalized cross-Kerr nonlinearities not only may give rise to the enhancement of one- and two-photon blockades as well as photon induced tunneling but can also provide more controllability over them. Furthermore, it is shown that in the regime of zero optomechanical coupling, with the aid of generalized cross-Kerr nonlinearity, one can generate multicomponent mechanical superposition states which exhibit robustness against system dissipations. We also study the steady-state entanglement between the microwave and mechanical modes, the results of which signify the role of generalized cross-Kerr nonlinearity in enhancing the entanglement in the regime of large-red detuning. The proposed generalized cross-Kerr optomechanical system can find potential applications in microwave quantum sensing, quantum telecommunication, and quantum information protocols.
\end{abstract}

\maketitle

\section{Introduction}

The past decade has witnessed significant theoretical as well as experimental progress in the field of cavity quantum optomechanics \cite{aspelmeyer2014,milburnbook} which explores the parametric coupling between the position of a mechanical oscillator and the frequency of a photon field inside a high-finesse optical (or microwave) cavity via the radiation pressure force. In a typical cavity optomechanical system the position of the mechanical oscillator modulates the resonance frequency of the cavity and gives rise to a nonlinear coupling between the cavity and the mechanical modes. Such a nonlinear coupling brings about broad applications ranging from addressing fundamental questions at the forefront of physics to realizing quantum technological goals. 
To cite some important examples, one could highlight exploring possible quantum gravitational phenomena \cite{pikovski2012}, Bell test for macroscopic mechanical entanglement \cite{vivoli2016,marinkovic2018,teleportation}, testing of the quantum-classical boundary \cite{isart2011}, high-precision position, mass, magnetic, or force sensing \cite{krause2012,chaste2012,motazediAVS2021,allahverdi,tsang2010,motazedifard2019,motazedifard2016,lepinay2021,ebrahimi2020,moller2017,bemani2021}, coherent photon-phonon conversion \cite{verhagen2012}, realization of the optomechanically induced transparency \cite{weis2010,aliGreen2021,vitaliOMIT,marquardtOMIT}, quantum state transfer \cite{palomaki2013}, optomechanical entanglement generation \cite{palomaki2013-2,barzanjehentanglement,dalafiQOC}, quantum information processing \cite{fiore2011}, generating nonclassical states of light \cite{brooks2012,aliDCEsqueezing2019} and mechanical motion \cite{wollman2015}, quantum correlations \cite{correlation,bemani2017}, and proposals for realizing the parametric  dynamical Casimir effect \cite{stefano2019,motazedifard2018,motazedifard2015,vincenzo2017}.

To observe and control quantum behaviors in a standard optomechanical system, it is essential to cool down the mechanical oscillator as close as possible to its motional ground state. To achieve this goal, a variety of methods have been proposed and realized \cite{schliesser2006,schliesser2008,xu2017,rossi2018,sommer2020}.
Even so, the ground-state cooling of a mechanical oscillator alone is not sufficient to observe quantum behavior. The other requirement is the necessity of being in the strong coupling regime \cite{groblacher2009} in which the coupling strength between the cavity and the mechanical modes is larger than the damping rates of  the cavity field and the mechanical oscillator. Nevertheless, the optomechanical coupling is usually weak, so the nontrivial quantum phenomena cannot be observed in the single-photon optomechanical coupling regime \cite{verhagen2012}.
To overcome this restriction, a strong external driving field can be applied to the cavity mode, and consequently a multiphoton strong-coupling regime is reached \cite{groblacher2009}, but with the sacrifice of the radiation-pressure nonlinearity. In addition, it has been shown \cite{gong2009,aldana2013} that under sufficiently strong driving, an optomechanical cavity behaves effectively as a rigid cavity filled with a nonlinear optical Kerr medium. Motivated by such an analogy between quantum optomechanics and nonlinear optics, the Kerr medium inside an optomechanical cavity has been proposed \cite{kumar2010} not only to enhance the optomechanical coupling strength, but also to avoid losing the nonlinearity.

Optomechanical systems coupled to a nonlinear inductive element (single-Cooper-pair transistor) \cite{heikkila2014} or a quantum two-level system (qubit) \cite{pirkkalainen2015} have also been proposed and experimentally studied as a promising platform for strengthening the single-photon radiation-pressure coupling by several orders of magnitude (about four to six orders of magnitude for typical experimental parameters). Moreover, it has been shown \cite{heikkila2014} that in such hybrid optomechanical systems 
a strong controllable cross-Kerr (CK) type of coupling $g_{\rm CK}\hat{n}_a\hat{n}_b$ between the cavity field and the mechanical oscillator, with respective number operators $\hat{n}_a$ and $\hat{n}_b$, can be realized.
The CK nonlinearity gives rise to a dispersive frequency shift in the mechanical (optical) mode with linear dependence versus the photon (phonon) number, as well as an optimal cooling or heating of the mechanical oscillator \cite{khan2015}. It can also significantly  enhance the quantum correlation between the optical and the mechanical modes \cite{chakraborty2017}.
Furthermore, it has been shown \cite{xiong2016} that in an optomechanical system with CK nonlinearity the optical bistability can be turned into tristable behavior. In Ref.\cite{zhang2017} the effects of the CK nonlinearity on the normal mode splitting, ground-state cooling, and steady-state entanglement in an optomechanical cavity assisted by an optical parametric amplifier have been studied. 
The CK effects on the optomechanically induced transparency phenomenon in a parity-time symmetric optomechanical system have also been explored \cite{zhao2018}.
Recently, the influence of the CK interaction on the few-photon optomechanical effects including the photon blockade (PB) and the mechanical cat state generation in a superconducting quantum circuit proposed in Ref.~\cite{heikkila2014} has been investigated \cite{zou2019}.
In addition, some realizable schemes have recently been proposed for boosting the photon-phonon CK coupling in optomechanical systems based on various methods such as two-photon parametric driving \cite{yin2018}, strong mechanical driving \cite{liao2020}, periodic modulation of the mechanical spring constant \cite{feng2021}, and utilization of Josephson (quantum) capacitance of a Cooper-pair box \cite{manninen2022}.

In this paper, inspired by the role of the nonlinear photon-phonon interaction in few-photon optomechanical phenomena, we introduce and investigate the influence of a higher-order nonlinear CK coupling, namely generalized CK nonlinearity, on photon blockade, mechanical Schrödinger cat state generation, and photon-phonon entanglement. This type of CK nonlinearity which is linearly related to the cavity photon number while depending quadratically on the mechanical phonon number (i.e., proportional to $\hat{n}_a\hat{n}_b^2$), can be realized in a tripartite  microwave optomechanical system composed of a single-Cooper-pair transistor, a microwave \textit{LC} resonator, and a micromechanical resonator. By calculating the steady-state equal-time second- and third-order correlation functions of the cavity mode we analyze, both analytically and numerically, the effect of generalized CK nonlinearity on the quantum statistics of intracavity photons including PB and photon induced tunneling (PIT). In particular, we show that photon bunching and antibunching can be effectively controlled by adjusting the system parameters. With regard to the generation on the mechanical cat state, we find based on both analytical and numerical solutions that in the regime of zero optomechanical coupling multi-component mechanical superposition states can be generated making use of the generalized CK nonlinearity. We also examine the robustness of these generated states against the system dissipations. Furthermore, our results reveal that although the generalized CK nonlinearity is weaker than the optomechanical and CK nonlinearities, it can increase the photon-phonon entanglement in the regime of large red-detuning.

The remainder of the paper is organized as follows. In Sec.~\ref{section:2} we introduce the physical system and derive an effective Hamiltonian containing the radiation pressure, the CK, and the generalized CK couplings. In Sec.~\ref{section:3} we analyze the PB and PIT phenomena in the system under consideration. Section~\ref{section:4} provides the discussion of the generation of Schrödinger cat sates in the mechanical mode. The steady-state entanglement between the microwave and mechanical modes is investigated in Sec.~\ref{section:5}. We present in Sec.~\ref{section:6} a brief discussion of the values of the experimental parameters required for implementation of the model. We summarize our conclusions in Sec.~\ref{section:7}.  The details of some derivations are contained in Appendixes~\ref{appendix1}-\ref{appendix4}.

\section{System Hamiltonian}\label{section:2}
\begin{figure}[t]
	\centering
	\includegraphics[width=8.6cm]{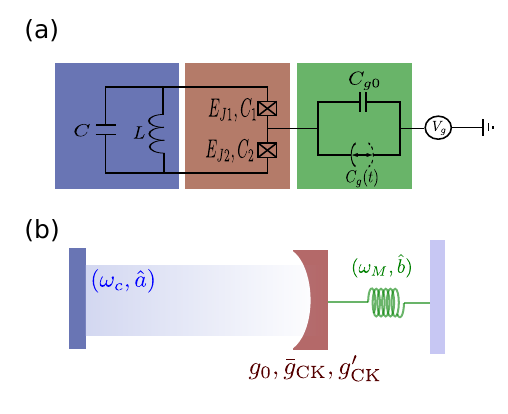}
	\caption{ (a) Schematic circuit diagram of the considered tripartite microwave optomechanical system \cite{heikkila2014} composed of a single-Cooper-pair transistor with Josephson energies $E_{J1,J2}$ and Josephson capacitances $C_{1,2}$ (the brown box), a microwave LC resonator (the blue box), and a micromechanical resonator with gate capacitance $C_{g0}$ which couples via a time-dependent capacitance $C_g(t)\equiv C_{g}[x(t)] \equiv C_g $ (the green box). Here, we consider the gate capacitor can be vibrated through modulating the movable part of the gate capacitance. (b) Equivalent cavity optomechanical system where the cavity mode $\hat{a}$ with frequency $\omega_c$ is coupled to the mechanical mode $\hat{b}$ with frequency $\omega_M$ through the radiation-pressure, the CK, and an additional higher order CK- types of interaction with coupling strengths $g_0,\bar{g}_{\rm CK}$, and $g'_{\rm CK}$, respectively  (see the text for details).
	}
	\label{fig1}
\end{figure}
As schematically shown in Fig.~\ref{fig1}(a), we consider a microwave optomechanical circuit which was proposed in Ref.~\cite{heikkila2014} for boosting the optomechanical radiation-pressure coupling. 
It consists of a single-Cooper-pair transistor with Josephson energies $E_{J1,J2}$ and Josephson capacitances $C_{1,2}$ (the brown box), a microwave \textit{LC} resonator (the blue box), and a micromechanical resonator with gate capacitance $C_{g0}$ which couples via a time-dependent capacitance $C_{g}(t)$ (the green-colored box). In this system, the nonlinearity of the Josephson effect can be exploited to achieve a strong radiation-pressure-type coupling between the mechanical and microwave modes \cite{heikkila2014}.

The total Hamiltonian describing the tripartite quantum system depicted in Fig.~\ref{fig1}(a) for the case of equal Josephson couplings $E_{J1}=E_{J2}=E_J$ can be written as \cite{heikkila2014}(see appendix \ref{appendix1}) 
\begin{gather}
\hat{H}_t=  \hbar \omega_c^0 \hat{c}^\dagger\hat{c} + \hbar \omega_M^0\hat{d}^\dagger\hat{d}+\frac{B_1}{2} \hat{\sigma}_1 + \frac{B_3}{2} \hat{\sigma}_3
\nonumber
\\
+ g_m \hat{\sigma_3} \hat{x}_m +
g_{q} \hat{\sigma_1}\hat{x}_c^2,
\label{H1}
\end{gather}
where $\omega_c^0=1/\sqrt{LC}$ and $\omega_M^0$ are the natural frequencies of the cavity and the mechanical modes, described by the annihilation operators $\hat{c}$ and $\hat{d}$, respectively.
In addition, $B_1$ and $B_3$ are the effective magnetic fields, $\hat{\sigma}_1$ and $\hat{\sigma}_3$ are Pauli matrices, $g_m$ and $g_q$ denote the coupling strengths between the Josephson junction qubit and two oscillators, and $\hat{x}_m=\hat{d}+\hat{d}^\dagger$ and $\hat{x}_c=\hat{c}+\hat{c}^\dagger$ stand for the position operators of the mechanical and cavity modes, respectively.

The Hamiltonian \eqref{H1} acts on two different subspaces; the qubit and the cavity and mechanical oscillator subspaces which represent the high-energy and low-energy subspaces, respectively.
We are interested in obtaining an effective low-energy Hamiltonian of the system. For this purpose, the Schrieffer-Wolff transformation \cite{bravyi2011}
can be utilized to decouple the high-energy and low-energy subspaces. However, in the dispersive regime where $\hbar\omega_{c,M}^0 \ll \abs{B}=\sqrt{B_1^2+B_3^2}$, and when all couplings are small it is sufficient to diagonalize the interaction part of the Hamiltonian \eqref{H1} in the qubit basis to obtain an effective low-energy Hamiltonian. Assuming the qubit to be in its ground state, we find $(\hbar=1)$ (see Appendix \ref{appendix2-2})  
\begin{eqnarray}\label{H_total}
&& \hat{H} =\hat{H}_0 + \hat{H}_{\rm op}+\hat{H}_{\rm CK}+\hat{H}_{\rm CK}',
\end{eqnarray}
with
\begin{eqnarray}
&& \hat{H}_0=\omega_c \hat{a}^\dagger \hat{a}+\omega_M\hat{b}^\dagger \hat{b},  \label{H_free} \\
&& \hat{H}_{\rm op}=g_0\hat{a}^\dagger \hat{a} (\hat{b}^\dagger+\hat{b}), \label{H_rp} \\
&& \hat{H}_{\rm CK}= \bar g_{\rm CK} \hat{a}^\dagger\hat{a}\hat{b}^\dagger\hat{b},  \label{H_ck} \\
&& \hat{H}_{\rm CK}'= g_{\rm CK}' \hat{a}^\dagger\hat{a} (\hat{b}^{\dagger}\hat{b})^2,  \label{H_c}
\end{eqnarray}
which describes an equivalent optomechanical cavity system where the cavity field interacts with the mechanical oscillator via the radiation-pressure, the CK, and an additional higher-order CK types of coupling [Fig. \ref{fig1}(b)].
Here $\hat{H}_0$ denotes the free energy of the cavity and mechanical modes where $\hat{a}(\hat{a}^\dagger)$ and $\hat{b}(\hat{b}^\dagger)$ are the annihilation (creation) operators of the cavity field and the mechanical mode, with effective resonant frequencies $\omega_c$ and $\omega_M$, respectively.
The Hamiltonian $\hat{H}_{\rm op}$ refers to the optomechanical radiation-pressure interaction with coupling strength $g_0$, which is four to six orders of magnitude larger than its counterpart in other conventional optomechanical systems \cite{heikkila2014}. The $\hat{H}_{\rm CK}$ originating from a quadratic-quadratic coupling term between $\hat{x}_c$ and $\hat{x}_m$ (see Appendix \ref{appendix2-2}), describes the CK interaction (linear phonon-number-dependent dispersive shift) between the cavity field and the mechanical mode, with the modified coupling strength $\bar{g}_{\rm CK}=g_{\rm CK}+g_{\rm CK}'$.
Finally, $\hat{H}_{\rm CK}'$ represents the higher-order CK interaction (quadratic phonon-number-dependent dispersive shift) with the coupling strength $g_{\rm CK}'$ which originates from an interaction that is quadratic with respect to $\hat{x}_c$ while being quartic in terms of $\hat{x}_m$ (see Appendix \ref{appendix2-2}). 

\begin{figure}[t]
	\centering
	\includegraphics[width=8.6cm]{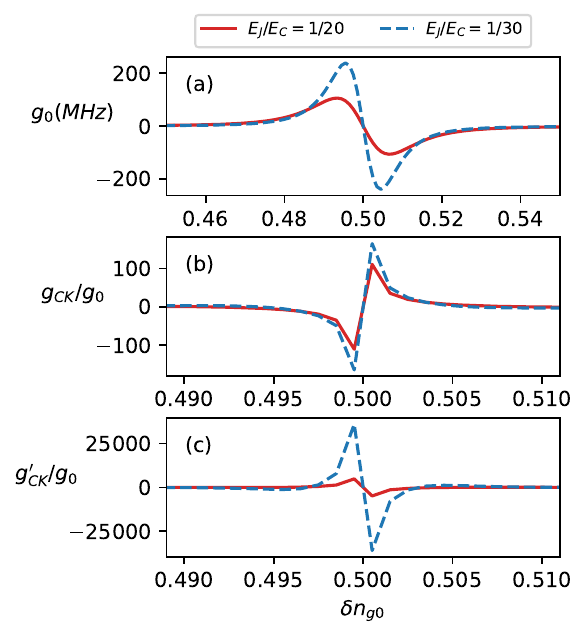}
	\caption{(Color online). Variation of (a) radiation pressure coupling, (b) CK coupling, and (c) the generalized CK coupling for different values of $E_J/E_C$ as a function of $\delta n_{g0}$. Here, $E_J/\hbar=10$ GHz.}
	\label{fig2}
\end{figure}

As shown in Appendix \ref{appendix2-2}, the coupling strengths $g_0,g_{\rm CK}$, and $g_{\rm CK}'$ in the Hamiltonian of Eq.~\eqref{H_total} depend on the ratio $E_J/E_C$ as well as $\delta n_{g0}$. Here $E_C=e^2/[2(C_1+C_2+C_{g0})]$ is the charging energy of the qubit and $\delta n_{g0}$ denotes the deviation from the two lowest charge states $\ket{\text{int}(n_{go})}=\ket{0}$ and $\ket{\text{int}(n_{go})+1}=\ket{1}$, with $n_{g0}=V_gC_{g0}/2e$ being the gate charge by which the energy difference of having zero or one Cooper pairs on the island can be tuned. Figure \eqref{fig2} shows the coupling strengths $g_0,g_{\rm CK}$, and $g_{\rm CK}'$ versus $\delta n_{g0}$ ($ \delta n_{g0}=[n_{g_0}-{\rm int}(n_{g_0})] \in [0,1] $) for different values of $E_J/E_C$.
As is seen, the coupling strengths reach their maximum values very close to the charge degeneracy point $\delta n_{g0}=0.5$, such that the maximum values are enhanced with deceasing the $E_J/E_C$ ratio. As explained in Appendix \ref{appendix2-2}, the Hamiltonian \eqref{H_total} is derived under certain limits (see Eqs.~\eqref{condition5} and~\eqref{general-condition}) which restrict the range of values the control parameters $E_J/E_C$ and $\delta n_{g0}$ can take.

To see how the CK and generalized CK couplings can influence the eigenenergy spectrum of the system, we need to diagonalize the Hamiltonian~\eqref{H_total}.
The energy eigenvalues associated with photons are explicit. However, the phononic part is complicated. 
It is straightforward to approximately diagonalize the phononic part of Hamiltonian $\hat{H}$ by using the well-known polaron transformation \cite{zou2019} as
\begin{equation}\label{polaron}
\hat{U}_P=\exp[f(\hat{n})(\hat{b}^\dagger-\hat{b})]=\exp(\frac{-g_0 \hat{n}}{\omega_M+\hat{n}\bar{g}_{\rm CK}}(\hat{b}^\dagger-\hat{b})),
\end{equation}
with $\hat{n}=\hat{a}^\dagger\hat{a}$, which displaces the resonator modes. The transformed Hamiltonian of the system reads
\begin{align}\label{H_Polaron2}
\hat{H}_P:=&\hat{U}^\dagger_P\hat{H}\hat{U}_P=
\hat{U}^\dagger_P[
\omega_c\hat{a}^\dagger\hat{a}+\omega_M\hat{b^\dagger}\hat{b}+g_0\hat{a}^\dagger\hat{a}(\hat{b}^\dagger+\hat{b})
\nonumber
\\
&+\bar g_{\rm CK} \hat{a}^\dagger\hat{a}\hat{b}^\dagger\hat{b}
]\hat{U}_P 
+\hat{U}^\dagger_P( g_{\rm CK}'\hat{a}^\dagger\hat{a}(\hat{b}^{\dagger}\hat{b})^2)\hat{U}_P 
\nonumber
\\
&=\omega_c\hat{n}+(\omega_M+ \bar g_{\rm CK}\hat{n})\hat{b}^\dagger \hat{b}-\frac{g_0^2}{\omega_M+ \bar g_{\rm CK}\hat{n}}\hat{n}^2
\nonumber
\\
&+g_{\rm CK}'\hat{U}^\dagger_P(\hat{n}
(\hat{b}^\dagger\hat{b})^2)\hat{U}_P.
\end{align}
Under the condition $\abs{f(n)}<1$, i.e., 
\begin{equation}\label{fn}
-1<\frac{g_0 n/\omega_M}{1+n\bar{g}_{\rm CK}/\omega_M}<1,
\end{equation} 
we  have $\hat{U}^\dagger_P[\hat{n}
(\hat{b}^\dagger\hat{b})^2]\hat{U}_P \approx \hat{n}
(\hat{b}^\dagger\hat{b})^2$, and thus the transformed Hamiltonian~\eqref{H_Polaron2} reduces to
\begin{align}\label{H_Polaron}
\hat{H}_P\approx&\omega_c\hat{n}+(\omega_M+ \bar g_{\rm CK}\hat{n})\hat{b}^\dagger \hat{b}
\nonumber
\\
&-\frac{g_0^2}{\omega_M+ \bar g_{\rm CK}\hat{n}}\hat{n}^2+g_{\rm CK}'\hat{n}
(\hat{b}^\dagger\hat{b})^2,
\end{align}

\begin{figure}[t]
	\centering
	\includegraphics[width=8.6cm]{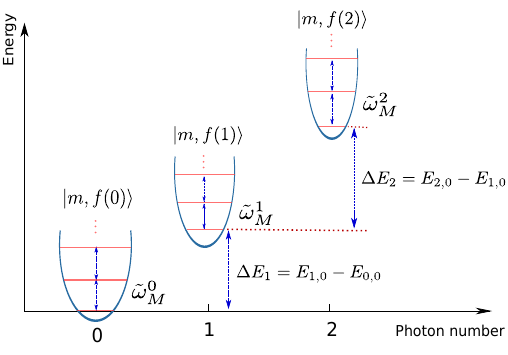}
	\caption{The qualitative (unscaled) diagram of the eigenenergy of the Hamiltonian $\hat{H}$. The horizontal axis shows photon subspace associated with $n=0,1,2$.}
	\label{fig3}
\end{figure}

Note that the condition~\eqref{fn} is justified in the regimes we will consider in Sec.~\ref{section:3}.
It is noteworthy that the radiation pressure vanishes ($g_0=0$) when the gate charge is tuned to the degeneracy point $\delta n_{g0}=0.5$, as Fig.~\ref{fig2}(a) shows. Consequently, in this case, the phononic part of the Hamiltonian~\eqref{H_total} is itself diagonal and the polaron transformation operator defined by Eq.~\eqref{polaron} reduces to the identity operator. As discussed in Sec.~\ref{section:4}, the possibility of achieving zero single-photon optomechanical coupling through adjusting the system parameters is a feature that can be used to generate the mechanical Schrödinger cat state in the system under consideration. The eigenstates and eigenenergies of Hamiltonian \eqref{H_Polaron} are given by   
\begin{equation}\label{eigen}
\hat{H}_P\ket{n}_a\ket{m}_b=E_{n,m}\ket{n}_a\ket{m}_b,
\end{equation}
where $\ket{n}_a$ and $\ket{m}_b$ are photonic and phononic number states, respectively, and
\begin{align}\label{energy}
E_{n,m}=n\omega_c+(\omega_M+ n \bar g_{\rm CK})m-\frac{(g_0 n)^2}{\omega_M+n \bar g_{\rm CK}}
+ng_{\rm CK}'m^2.
\end{align}
The eigenvalues of the Hamiltonian~\eqref{H_total} are the same as $E_{n,m}$, but its eigenstates differ from those of $\hat{H}_P$:
\begin{equation}\label{eigen-origin}
\hat{H}\ket{n}_a\ket{m,f(n)}_b=E_{n,m}\ket{n}_a\ket{m,f(n)}_b,
\end{equation}
where $\ket{m,f(n)}_b$ is the displaced number state defined as \cite{oliveira1990,nieto1997}:
\begin{equation}
\ket{m,f(n)}_b:=\hat{U}_P\ket{m}_b= \exp(f(n)(\hat{b}^\dagger-\hat{b}))\ket{m}_b.
\end{equation}
with
\begin{equation}
f(n)=\frac{-g_0 n}{\omega_M+n \bar g_{\rm CK}}.
\end{equation}
Note that the set of displaced number states is a complete set for an arbitrary value of $f(n)$.
We define the generalized shifted mechanical frequency as
\begin{equation}\label{modefiy-omega}
\tilde{\omega}_M^n= \omega_M + n(\bar g_{\rm CK} + g_{\rm CK}'),
\end{equation}
where $n$ is the photon number.
The diagram of the eigenenergy of the Hamiltonian $\hat{H}$ is depicted in Fig.~\eqref{fig3}. According to Eqs.~\eqref{energy} and \eqref{modefiy-omega}, the different values of photon and phonon energy levels depend on the CK and generalized CK nonlinear terms, giving rise to an anharmonicity to the energy-level structure. This anharmonicity causes the symmetry-breaking behavior of the system eigenenergy spectrum which is responsible for the phenomenon of blockade in the system.

\section{photon blockade and photon-induced tunneling}\label{section:3}
Using the Hamiltonian~\eqref{H_total}, we now investigate photon statistical properties including PB and PIT in the system under consideration. 
To this end, we assume that the cavity is weakly driven by a monochromatic laser field. Then, the Hamiltonian of the total system reads
\begin{align}
\hat{H}_{\rm sys}=&\omega_c \hat{a}^\dagger \hat{a}+\omega_{M}\hat{b}^\dagger \hat{b}+g_0 \hat{a}^\dagger \hat{a} (\hat{b}^\dagger+\hat{b})
+\bar g_{\rm CK} \hat{a}^\dagger \hat{a} \hat{b}^\dagger\hat{b}
\nonumber
\\
& + g_{\rm CK}' \hat{a}^\dagger\hat{a}(\hat{b}^\dagger\hat{b})^2
+\Omega(\hat{a}e^{i\omega_d t}+\hat{a}^\dagger e^{-i\omega_d t}),
\end{align}
where $\Omega=\sqrt{2\kappa \mathcal{P}/\hbar \omega_d}$ denotes the amplitude of the driving field with power $\mathcal{P}$ and frequency $\omega_d$, and $\kappa$ is the cavity decay rate.

In the frame rotating at the driving laser frequency  $\omega_d$, the Hamiltonian of the system can be written as
\begin{align}\label{H-driving}
\hat{\tilde{H}}_{\rm sys}= &\Delta_c \hat{a}^\dagger \hat{a}+\omega_{M}\hat{b}^\dagger \hat{b}+g_0 \hat{a}^\dagger \hat{a} (\hat{b}^\dagger+\hat{b})
\nonumber
\\
& +\bar g_{\rm CK} \hat{a}^\dagger \hat{a} \hat{b}^\dagger \hat{b}+ g_{\rm CK}' \hat{a}^\dagger\hat{a}(\hat{b}^\dagger\hat{b})^2
+\Omega(\hat{a}+\hat{a}^\dagger),
\end{align}
where $\Delta_c=\omega_c-\omega_d$ is the detuning between the cavity and driving field. 

To quantitatively study the quantum statistics of intracavity photons, we consider the steady-state equal-time second-and third-order correlation functions of the cavity mode defined, respectively, as
\begin{subequations}\label{correlation-full}
	\begin{eqnarray}
	g^{(2)}{(0)} = \frac{ \langle \hat a^{\dag 2} \hat a^2  \rangle_{ss}}{\langle \hat a^\dag \hat a  \rangle_{ss}^2}, 
	\end{eqnarray}
	\begin{eqnarray}
	g^{(3)}{(0)} =\frac{ \langle \hat a^{\dag 3} \hat a^3  \rangle_{ss}}{\langle \hat a^\dag \hat a \rangle_{ss}^3}, \qquad
	\end{eqnarray}
\end{subequations}
where the average values are taken over the steady state of the system. Typically, $g^{(2)}(0)<1$ corresponds to the sub-Poissonian photon statistics, which is a nonclassical effect often referred to as photon anti-bunching effect.
Specifically, $g^{(2)}(0)\rightarrow 0$ is a signature of the complete single-photon blockade (1PB) effect, in which only one photon can be excited in the cavity mode, i.e., an ideal single-photon source. The two-photon blockade (2PB) effect occurs when $g^{(2)}(0) \geq 1$ and $g^{(3)}(0)<1$ \cite{hamsen2017, huang2018}.
As the name implies, the 2PB means that the generation of the second photon will block the emergence of the third photon, i.e., two-photon bunching and three photon antibunching. On the other hand, $g^{(n)}(0)>1 (n=2,3)$ indicates the photon bunching effect, in which the excitation of the first photon contributes to the excitation of the second or third photons, and the photons exhibit the super-Poissonian statistics. This phenomenon which is referred to as two photon-induced tunneling (2PIT) or three photon-induced tunneling (3PIT) depending on $1<g^{(3)}(0)<g^{(2)}(0)$ or $1<g^{(2)}(0)<g^{(3)}(0)$, has been explored theoretically \cite{xu2013,huang2018,xie2016,laszyk2019} and observed experimentally \cite{faraon2008,majumdar2012}.

In the following, we calculate the equal-time second-and third-order correlation functions of the cavity mode in the optomechanical system under consideration through analytically solving the non-Hermitian Schrödinger equation and numerically simulating the quantum master equation.

\subsection{Approximate analytical solution}
As long as the driving laser field is weak enough, i.e., $\Omega\ll~\kappa$, only the lower energy levels of the cavity modes are occupied. In this case, the driving term in the Hamiltonian \eqref{H-driving} can be treated as a perturbation. Truncating the Hilbert space of the cavity field up to $n=3$, a general state of the system in this subspace can be expressed as
\begin{equation} \label{expansion}
\ket{\psi(t)}=\sum_{m=0}^{\infty}\sum_{n=0}^{3}C_{n,m}(t)\ket{n}_a\ket{m,f(n)}_b,
\end{equation}
where the coefficient $C_{n,m}(t)$ stands for the probability amplitude of the corresponding state $\ket{n}_a\ket{m,f(n)}_b$.
We phenomenologically add an anti-Hermitian term to the Hamiltonian, given in Eq.~\eqref{H-driving} \cite{liao2013,plenio1998} to take into account the dissipation of the cavity mode. Considering the case of zero-temperature photon bath, the effective non-Hermitian Hamiltonian takes the form
\begin{equation}\label{H-eff}
\hat{H}_{\text{eff}}=\hat{\tilde{H}}_{\rm sys}-i\frac{\kappa}{2}\hat{a}^\dagger\hat{a}.
\end{equation}
It should be noted that since the cavity decay dominates over the mechanical dissipation, here we approximately neglect the dissipation of the phononic mode which is justified in the time scale $1/\kappa\ll t\ll 1/\gamma$ where $\gamma$ represents the mechanical decay rate. However, we will later take into account the mechanical dissipation in our numerical calculation.

Inserting  the non-Hermitian Hamiltonian \eqref{H-eff} and the general state vector \eqref{expansion} into the Schrödinger equation $i\frac{d}{dt}\ket{\psi(t)}=\hat{H}_{\rm eff}\ket{\psi(t)}$, we obtain a set of linear differential equations for the probability amplitudes, which reads
\begin{subequations}\label{cdot}
\begin{eqnarray}
\dot{C}_{0,m}=&& -i \tilde{E}_{0,m} C_{0,m}- i\Omega\sum_{l=0}^{\infty} ~_b\!\braket{m, f(0)}{l, f(1)}_b C_{1,l},
\\
\dot{C}_{1,m}=&&-(i \tilde{E}_{1,m}+\kappa/2)C_{1,m}-i\Omega\sum_{l=0}^{\infty} ~_b\!\braket{m,f(1)}{l,f(0)}_b C_{0,l}
\nonumber
\\
&&-i\sqrt{2}\Omega \sum_{l=0}^{\infty} ~_b\!\braket{m, f(1)}{l.f(2)}_b C_{2,l},
\\
\dot{C}_{2,m}=&& -(i \tilde{E}_{2,m}+\kappa)C_{2,n}-i\sqrt{2}\Omega\sum_{l=0}^{\infty} ~_b\!\braket{m,f(2)}{l,f(1)}_b C_{1,l}
\nonumber
\\
&&-i\Omega\sqrt{3}\sum_{l=0}^{\infty} ~_b\!\braket{m,f(2)}{l,f(3)}_b C_{3,l},
\\
\dot{C}_{3,m}=&& -(i\tilde{E}_{3,m}+3\kappa/2)C_{3,m}
\nonumber
\\
&&-i\Omega\sqrt{3}\sum_{l=0}^{\infty} ~_b\!\braket{m,f(3)}{l,f(2)}_b C_{2,l}.
\end{eqnarray}
\end{subequations}
In these equations, the quantities $\tilde{E}_{nm}$ denote the eigenvalues of the Hamiltonian \eqref{H-driving} in the absence of the cavity field driving,
\begin{align}\label{energy-tilde}
\tilde{E}_{n,m}=n\Delta_c+(\omega_M+ n \bar g_{\rm CK})m-\frac{(g_0 n)^2}{\omega_M+n \bar g_{\rm CK}}
+g_{\rm CK}'nm^2.
\end{align}
Furthermore, the transition amplitudes $_b\braket{m,f(n)}{m',f(n')}_b$, called Franck-Condon factors \cite{franck1926,condon1926}, are determined by the relation 
\begin{equation}\label{Franck}
\prescript{}{b}{\braket{m,f(n)}{m',f(n')}_b} = \prescript{}{b}{\bra{m}}\exp[\Big\{f(n)-f(n')\Big\}(\hat{b}^\dagger-\hat{b})]\ket{m'}_b,
\end{equation}
where the matrix elements can be calculated based on the relation \cite{oliveira1990}
\begin{gather}\label{matrix_element}
\prescript{}{b}{\bra{m}\exp[x(\hat{b}^\dagger-\hat{b})]\ket{m^\prime}_b} =
\nonumber
\\
\begin{cases}
\sqrt{\frac{m!}{m'!}}e^{-x^2/2}(-x)^{m'-m}L_m^{m'-m}(x^2), \qquad m'\ge m,
\\
\sqrt{\frac{m'!}{m!}}e^{-x^2/2}(x)^{m-m'}L_{m'}^{m-m'}(x^2),\qquad \space m > m',
\end{cases}
\end{gather}
where $L_m^{m'}(x)$ is the associated Laguerre polynomial.

Under the weak-driving assumption, we have the approximate scales $C_{n,m}\approx(\Omega/\kappa)^n$ for $n=0,1,2,3$.
Thus, we can approximately solve the set of Eqs.~(\ref{cdot}a-d) by using a perturbation method by discarding higher-order terms in the equations motion for the lower-order variables. This approximation has been widely used in cavity QED \cite{leach2004,carmichael1991} and optomechanical systems \cite{liao2013,komar2013} for investigating the photon statistics. Assuming the cavity field is initially in the vacuum state, i.e., $C_{n,m}(0)=0$ for $n=1,2,3$, the long-time solutions of the equations of motion for the probability amplitudes are approximately given by  
\begin{subequations}\label{c}
	\begin{eqnarray}
	C_{0,m}(\infty)=&& C_{0,m}(0) e^{-i\tilde{E}_{0,m} t},
	\\
	C_{1,m}(\infty)=&&-\Omega \sum_{l=0}^{\infty}\frac{ ~_b\!\braket{m,f(1)}{l,f(0)}_bC_{0,l}(0)e^{-i\tilde{E}_{0,l} t}}{\tilde{E}_{1,m}-\tilde{E}_{0,l}-i\kappa/2},
	\\
	C_{2,m}(\infty)=&&\sqrt{2}\Omega^2 \sum_{l,k=0}^{\infty} \frac{~_b\!\braket{m,f(2)}{l,f(1)}_b ~_b\!\braket{l,f(1)}{k,f(0)}_b}{\tilde{E}_{2,m}-\tilde{E}_{0,k}-i\kappa}
	\nonumber
	\\
	&&\times \frac{C_{0,k}(0)e^{-i\tilde{E}_{0,k} t}}{\tilde{E}_{1,l}-\tilde{E}_{0,k}-i\kappa/2},
	\\
	C_{3,m}(\infty)=&&-\sqrt{6}\Omega^3 \sum_{l,k,j=0}^{\infty}\frac{~_b\!\braket{m,f(3)}{l,f(2)}_b}{\tilde{E}_{3,m}-\tilde{E}_{0,j}-i3\kappa/2}
	\nonumber
	\\
	&&\times \frac{~_b\!\braket{l,f(2)}{k,f(1)}_b ~_b\!\braket{k,f(1)}{j,f(0)}_b}{\tilde{E}_{2,l}-\tilde{E}_{0,j}-i\kappa}
	\nonumber
	\\
	&&\times \frac{C_{0,j}(0)e^{-i\tilde{E}_{0,j} t}}{\tilde{E}_{1,k}-\tilde{E}_{0,j}-i\kappa/2},
	\end{eqnarray}
\end{subequations}
where the initial probability amplitudes $C_{0,m}(0)$, $C_{0,l}(0)$, $C_{0,j}(0)$, and $C_{0,k}(0)$ are determined by the initial state of the mechanical mode. We assume that the mechanical oscillator is initially prepared in its ground state $\ket{0}_b$, i.e., $C_{0,m}(0)=\langle 0\vert m\rangle_b=\delta_{m,0}$ (note that $ \vert m,f(n=0)=0 \rangle_b= \ket{m}_b $).
Therefore, with the probability amplitudes given in Eqs.~(\ref{c}a-d) and $P_n=\sum_{m=0}^{\infty}\abs{C_{n,m}}^2$ for $n=1,2,3$, we obtain the single-photon, two-photon, and three-photon probabilities, respectively, as
\begin{gather}\label{p1}
P_1=\sum_{m=0}^{\infty}\Bigg|
\Omega\frac{~_b\!\braket{m,f(1)}{0}_b}{\tilde{E}_{1,m}-\frac{i}{2}\kappa}
\Bigg|^2,
\end{gather}
\begin{gather}\label{p2}
P_2=\sum_{m=0}^{\infty}\Bigg|
\sum_{l=0}^{\infty}\sqrt{2}\Omega^2\frac{~_b\!\braket{m,f(2)}{l,f(1)}_b}{\tilde{E}_{2,m}-i\kappa}
\cross \frac{~_b\!\braket{l,f(1)}{0}_b}{\tilde{E}_{1,l}-\frac{i}{2}\kappa}
\Bigg|^2,
\end{gather}
\begin{gather}
P_3=\sum_{m=0}^{\infty}\Bigg|
\sum_{l,k=0}^{\infty}\sqrt{6}\Omega^3\frac{~_b\!\braket{m,f(3)}{l,f(2)}_b}{\tilde{E}_{3,m}-i\frac{3}{2}\kappa} 
\times \frac{~_b\!\braket{l,f(2)}{k,f(1)}_b}{\tilde{E}_{2,l}-i\kappa}
\nonumber
\\
\times \frac{~_b\!\braket{k,f(1)}{0}_b}{\tilde{E}_{1,k}-\frac{i}{2}\kappa}\Bigg|^2.
\label{p3}
\end{gather}

Using Eq.~\eqref{expansion}, when the system is in the steady state, the equal-time second- and third-order correlation functions \eqref{correlation-full} can be written, respectively, as
\begin{subequations}\label{correlation}
	\begin{gather}
	g^{(2)}(0)=\frac{2P_2}{(P_1+2P_2)^2}\approx\frac{2P_2}{P_1^2},
	\label{correlation1}
	\\
	g^{(3)}(0)=\frac{6P_3}{(P_1+2P_2+3P_3)^3}\approx\frac{6P_3}{P_1^3},
	\label{correlation2}
	\end{gather}
\end{subequations}
where the photon probabilities $P_i (i=1,2,3)$ are given by Eqs.~\eqref{p1}-\eqref{p3}, and we have used $P_1 \gg P_2 \gg P_3$ under the weak-driving condition. After some straightforward calculations, we can approximate the correlation functions as
\begin{equation}\label{correlation_approximation-g2}
g^{(2)}(0)\approx4\Bigg|\frac{\tilde{E}_{1,0}-i\frac{\kappa}{2}}{\tilde{E}_{2,0}-i\kappa}\Bigg|^2
=\frac{4(\Delta_c-\delta^{[1]})^2+\kappa^2}{(2\Delta_c-\delta^{[2]})^2+\kappa^2} ,
\end{equation}
\begin{align}\label{correlation_approximation-g3}
g^{(3)}(0)&\approx 36\Bigg|\frac{(\tilde{E}_{1,0}-i\frac{\kappa}{2})^2}{\tilde{E}_{2,0}-i\kappa}
\times
\frac{1}{\tilde{E}_{3,0}-i\frac{3\kappa}{2}}\Bigg|^2
\nonumber
\\
&= \frac{144(\Delta_c-\delta^{(1)})^4+9\kappa^4-72\kappa^2(\Delta_c-\delta^{(1)})}{[(2\Delta_c-\delta^{(2)})^2+\kappa^2][4(3\Delta_c-\delta^{(3)})^2+9\kappa^2]},
\end{align}
where
\begin{equation}\label{delta}
\delta^{[n]}=\frac{g_0^2n^2}{\omega_M+n\bar g_{\rm CK}}, \qquad (n=1,2).
\end{equation}
denotes the $n$-photon frequency shift with $\bar{g}_{\rm CK}$ the modified cross-Kerr coupling.

The simplified analytic solution of $g^{(2)}(0)$, i.e., Eq.~\eqref{correlation_approximation-g2}, has the same form as that given in \cite{zou2019} except that the cross-Kerr coupling $g_{\rm CK}$ is replaced by the modified cross-Kerr coupling $\bar{g}_{\rm CK}=g_{\rm CK}+g_{\rm CK}'$. As can be seen in Eq.~\eqref{correlation_approximation-g2}, the value of $g^{(2)}(0)$ depends on $\delta^{[1]}$ and $\delta^{[2]}$.
For the single-photon resonance case, i.e., $\Delta_c=\delta^{[1]}$, the correlation function $g^{(2)}(0)$ takes the form $g^{(2)}_{\rm SPR}(0)=\frac{\kappa^2}{(2\delta^{[1]}-\delta^{[2]})^2+\kappa^2}$ which shows that the larger the difference $2\delta^{[1]}-\delta^{[2]}$ results in a more effective photon blockade $[g^{(2)}_{\rm SPR}(0) < 1]$.
Expanding $\delta^{[1]}$ and $\delta^{[2]}$ up to $\bar{g}_{\rm CK}/\omega_M$ we get $2\delta^{[1]}-\delta^{[2]}\approx(2g_0^2/\omega_M)[(3\bar{g}_{\rm CK}/\omega_M)-1]$.
Thus depending on whether  is negative or positive, the photon blockade is enhanced or weakened compared to the case when the quadratic phonon-number-dependent dispersive shift is not considered, $g_{\rm CK}'=0$ \cite{zou2019}.
On the other hand, for the two-photon resonant case, i.e., $2\Delta_c=\delta^{[2]}$, the second-order correlation function becomes $g^{(2)}_{\rm TPR}(0)=\frac{(2\delta^{[1]}-\delta^{[2]})^2+\kappa^2}{\kappa^2}>1$, which corresponds to the PIT.


\subsection{Numerical simulation}
In order to verify the validity of our previous analytical treatment, we now numerically simulate the photon statistical properties by employing the quantum master equation. Considering both optical and mechanical dissipations, the dynamical evolution of the system is described by the master equation of the density operator $\hat{\rho}$,
\begin{eqnarray}\label{master-equation}
\dot{\hat{\rho}}=&&i[\hat{\rho},\hat{H}_{\rm sys}]+\frac{\kappa}{2}(\hat{a}\hat{\rho}\hat{a}^\dagger-\hat{a}^\dagger\hat{a}\hat{\rho}-\hat{\rho}\hat{a}^\dagger\hat{a})
\nonumber 
\\
&&+\frac{\gamma}{2}(\bar{n}_{\rm th}+1)(\hat{b}\hat{\rho}\hat{b}^\dagger-\hat{b}^\dagger\hat{b}\hat{\rho}-\hat{\rho}\hat{b}^\dagger\hat{b})
\nonumber 
\\
&&+\frac{\gamma}{2}\bar{n}_{\rm th}(\hat{b}^\dagger\hat{\rho}\hat{b}-\hat{b}\hat{b}^\dagger\hat{\rho}-\hat{\rho}\hat{b}\hat{b}^\dagger).
\end{eqnarray}
where the Hamiltonian $\hat{H}_{\rm sys}$ is given in Eq.~\eqref{H-driving}, $\gamma$ denotes the mechanical decay rate, and $\bar{n}_{\rm th}=[\exp(\omega_M/\rm k_BT)-1]^{-1}$
is the average thermal phonon number of the mechanical oscillator at temperature $T$, with $k_B$ being the Boltzmann constant. Here, we assume that the cavity field is coupled to a vacuum reservoir.

By using the quantum toolbox in PYTHON (QUTIP) \cite{qutip1,qutip2}, we numerically solve the master equation to obtain the steady-state density matrix of the system  $\hat{\rho}_{\rm ss}$. 
Then, the exact numerical results for the equal-time second-order $(n=2)$ and third-order $(n=3)$ correlation functions can be obtained by 
\begin{equation}\label{correlation-numerical}
g^{(n)}(0)=\frac{\Tr(\hat{a}^{\dagger n}\hat{a}^n\hat{\rho}_{\rm ss})}{[\Tr(\hat{a}^\dagger\hat{a}\hat{\rho}_{\rm ss})]^n}, \quad n=2,3.
\end{equation}

\begin{figure}[t]
	\centering
	\includegraphics[width=8.6cm]{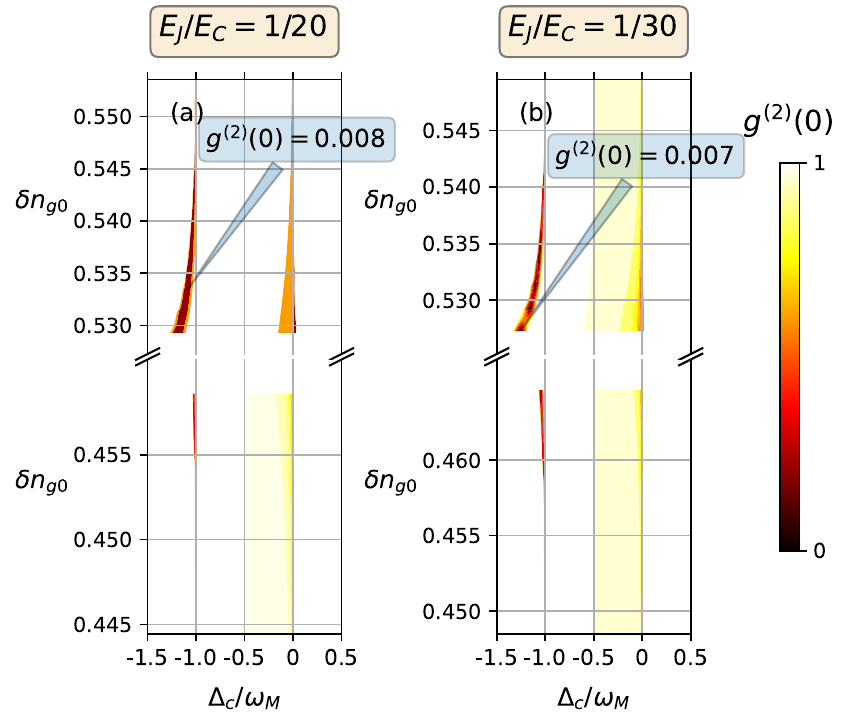}
	\caption{(Color online) Contour plots of the steady-state second-order correlation function $ g^{(2)}(0) $ versus the normalized cavity-laser detuning $\Delta_c /\omega_M$ and gate charge deviation $\delta n_{g0}$ for different coupling regimes $(a)$ $E_J/E_C=1/20$, and $(b)$ $E_J/E_C=1/30$. 
	The other system parameters are taken as  $\omega_c/2\pi=5$ GHz, $\omega_M/2\pi=10$ MHz, $\kappa/\omega_M=0.01$, $\gamma/\omega_M=0.001$, $\Omega/\omega_M=0.001$, and $\bar{n}_{\rm th}=0$.}
	\label{fig4}
\end{figure}

\begin{figure}[t]
	\centering
	\includegraphics[width=8.6cm]{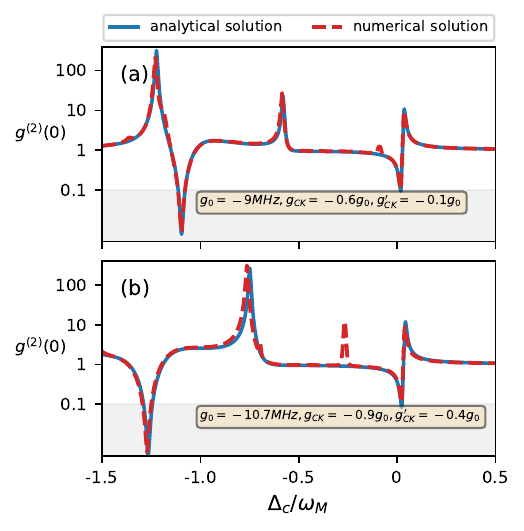}
	\caption{(Color online) Steady-state second-order correlation function, $ g^{(2)}(0) $, versus the  normalized cavity-laser detuning $\Delta_c/\omega_M$ for (a) $E_J/E_C = 1/20,\delta n_{g0}=0.533$, and (b) $E_J/E_C = 1/30,\delta n_{g0}=0.527$. The corresponding values of the coupling strengths are also given.
	The blue-solid and red-dashed lines are, respectively, referred to the analytical and numerical solutions.
	Other system parameters are the same as in Fig.~\eqref{fig4}.}
	\label{fig5}
\end{figure}

\begin{figure}[t]
	\centering
	\includegraphics[width=8.6cm]{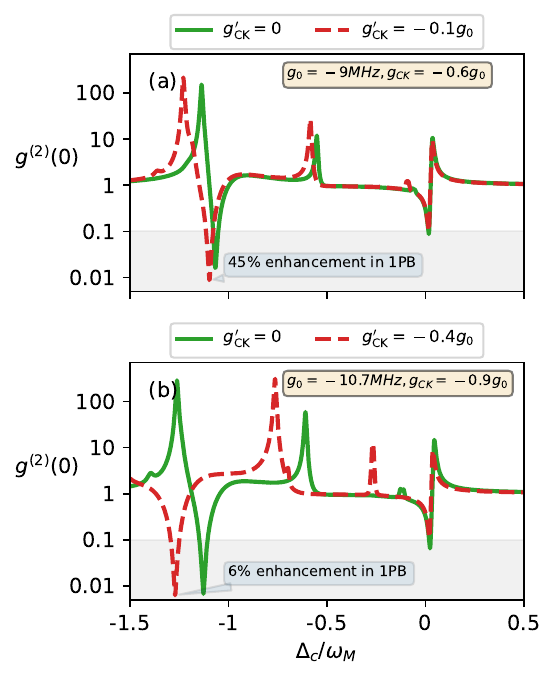}
	\caption{(Color online) Steady-state second-order correlation function, $ g^{(2)}(0) $, versus $\Delta_c/\omega_M$ with the same coupling strengths $g_0$ and $g_{\rm CK}$ in (a) and (b) as those chosen in Fig.~\ref{fig5}(a) and~\ref{fig5}(b), respectively.
	The green-solid and red-dashed lines are, respectively, referred to $g_{\rm CK}'=0$ and  $g_{\rm CK}'\ne0$.	 
	Other system parameters are the same as in Fig.~\eqref{fig4}.}
	\label{fig6}
\end{figure}


In order to verify our analysis, we plot in Fig.~\eqref{fig4} the equal-time second-order correlation function $g^{(2)}(0)$ at the steady state versus the normalized cavity-laser detuning $\Delta_c/\omega_M$ and gate charge deviation $\delta n_{g0}$, for the different coupling regimes $E_J/E_C=1/20$, and $E_J/E_C=1/30$ which imply different values of the coupling strengths $g_0,g_{\rm CK},g_{\rm CK}'$ through analytical calculation of Eq.~\eqref{correlation1}. 
In this figure, the range of the gate charge deviation $\delta n_{g0}$ is limited ( approximately $\delta n_{g0} \subset (0.450,0.460) \bigcup (0.530,0.560)$), since only in this range the general conditions Eqs.~\eqref{general-condition} and ~Eq.~\eqref{condition5} for obtaining Hamiltonian~\eqref{H_total}, and the particular limitation related to the polaron transformation given in Eq.~\eqref{fn} satisfied. 
To achieve precise compliance with Eq.~\eqref{general-condition} in numerical calculations, we impose the condition $G/g\ll0.1$, where $G=\{G_2, G_4\}$ and $g=\{g_0, g_{\rm CK}, g_{\rm CK}'\}$.
As can be seen, over a wide range $\Delta_c/\omega_M\subset [-0.5,0]$ weak PB happens. 
In this case, a relatively strong 1PB occurs near the blue detuning $\Delta_c\approx-\omega_M$; the minimum values of the second-order correlation function are shown in each plot. However, in the range $\Delta_c/\omega_M\subset (-1,-0.5) \bigcup (0,0.5)$, the cavity photons satisfy Poissonian or super-Poissonian statistics [$g^{(2)}(0)\ge1$]. 
Also, this figure reveals that the PB phenomenon is enhanced at upper values of the gate charge deviation.

As indicated in Fig.~\eqref{fig4}, the minimum values of the second-order correlation function for $E_J/E_C=1/20$ and $E_J/E_C=1/30$ are, respectively, $g^{(2)}(0)=0.008$ and $g^{(2)}(0)=0.007$ which occur at respective gate charge deviations of $\delta n_{g0}=0.533$ and $\delta n_{g0}=0.527$. 
Note that to have the desired optimum points of the deviation gate charge, $ \delta n_{g0} $, to obtain the desired values of these correlation functions, one must precisely control the change of the deviation gate charge and its fluctuations should be smaller $ 10^{-3}$. To achieve this, one needs to use the precise electronics elements which are usual in the advanced Labs [see the Sec.~(\ref{section:6})].

Figure \eqref{fig5} shows the second-order correlation function for these values of gate charge deviation.
The corresponding values of the coupling strengths ($g_0,g_{\rm CK},g_{\rm  CK}'$) are also given. 
The analytical and numerical results almost agree well with each other.
Moreover, each plot in Fig.~\eqref{fig5} reaches two peaks corresponding to the PIT effect in the system.

\begin{figure}[t]
	\centering
	\includegraphics[width=8.6cm]{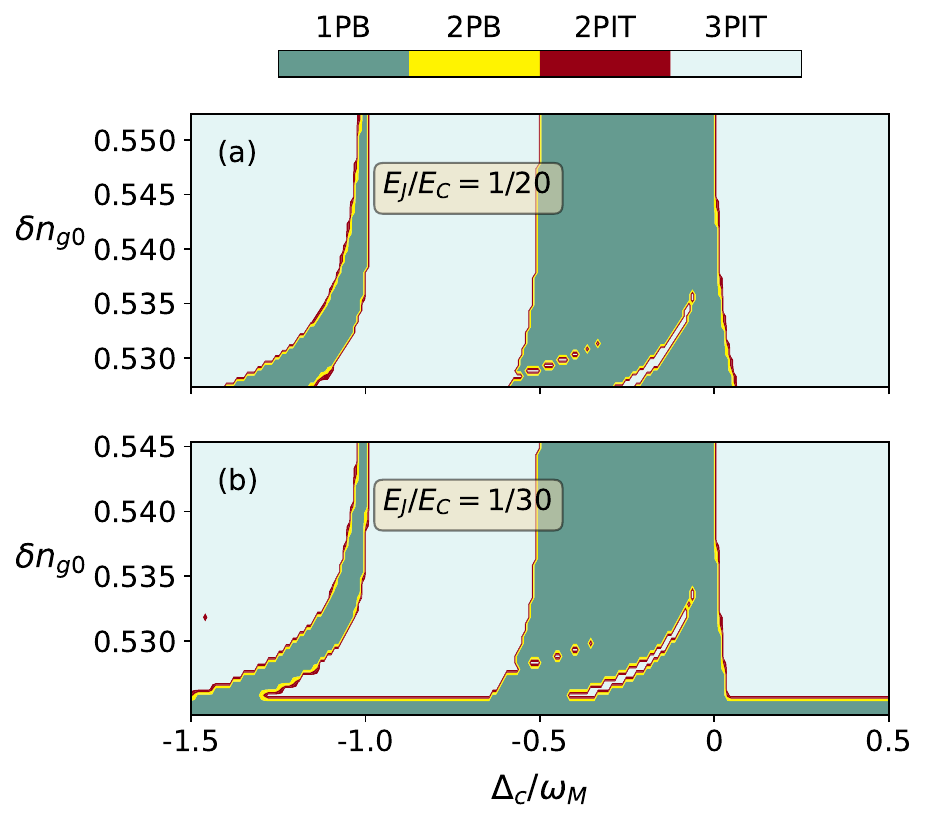}
	\caption{(Color online) The map of four effects including 1PB, 2PB, 2PIT, and 3PIT versus $\Delta_c/\omega_M$  and the gate charge deviation $\delta n_{g0}$, for different coupling regimes: $(a)$ $E_J/E_C=1/20$, and $(b)$ $E_J/E_C=1/30$.
	Other system parameters are the same as in Fig.~\eqref{fig4}.}
	\label{fig7}
\end{figure}

To illustrate more clearly the effect of second-order dispersive shift $g_{\rm CK}'$ on the photon statistics for the different values of the control parameters ($\delta n_{g0}, E_J/E_C$), we numerically plot in Fig.~\eqref{fig6} the steady-state correlation function $g^{(2)}(0)$ as a function of the normalized cavity-laser detuning $\Delta_c/\omega_M$ with $g_{\rm CK}'=0$ (green-solid line) and $g_{\rm CK}'\ne  0$ (red-dashed line) in each plot.
From this figure, we can see that the second-order dispersive shift can enhance the 1PB about 45\% near the blue detuning $\Delta_c=-1.1\omega_{M}$ in Fig.~\ref{fig6}(a) and also about 6\% near the blue detuning $\Delta_c=-1.1\omega_{M}$ in Fig.~\ref{fig6}(b) compared with the typical case $g_{\rm CK}'=0$.
Moreover, we find similar results for the peak of $g^{(2)}(0)$ in the plots. In conclusion,  the second-order dispersive shift can strengthen the PIT.
Furthermore, for improving the photon blockade effect of the system, it is essential to find appropriate relations among the coupling strengths $g_0,g_{\rm CK}$, and $g_{\rm  CK}'$. It should be noted that selecting higher values for the second-order dispersive shift is not the correct strategy to enhance the 1PB effect.
The steady-state second- and third-order correlation functions $g^{(2)}(0)$ and $g^{(3)}(0)$ are computed numerically to produce Figs.~\ref{fig7}(a) and~\ref{fig7}(b), which illustrate four distinct effects including 1PB, 2PB, 2PIT, and 3PIT as a function of the normalized cavity-laser detuning $\Delta_c/\omega_M$ and the gate charge deviation $\delta n_{g0}$.
While the general conditions Eq.~\eqref{general-condition} and Eq.~\eqref{condition5} limit the range of the gate charge deviation similar to Fig.~\eqref{fig4}, the particular condition Eq.~\eqref{fn} is not taken into account in Fig.~\eqref{fig7}, since the steady-state second- and third-order correlation functions $g^{(2)}(0)$ and $g^{(3)}(0)$ are calculated directly form Eq.~\eqref{H_total} by solving master equation numerically.
It can be observed that the range $\Delta_c/\omega_M \subset(-1.5,-1) \bigcup (-0.5,0)$ exhibits 1PB, while the range $\Delta_c/\omega_M \subset(-1,-0.5) \bigcup (0,0.5)$ is relevant to the 3PIT. 
The narrow boundary between the two previously mentioned areas (1PB, 3PIT), marked by 2PB and 2PIT, corresponds to the tiny ranges of $\Delta_c/\omega_M$ and $\delta n_{g0}$ in which the 2PB effect can occur in the system. 

\begin{figure}[t]
	\centering
	\includegraphics[width=8.6cm]{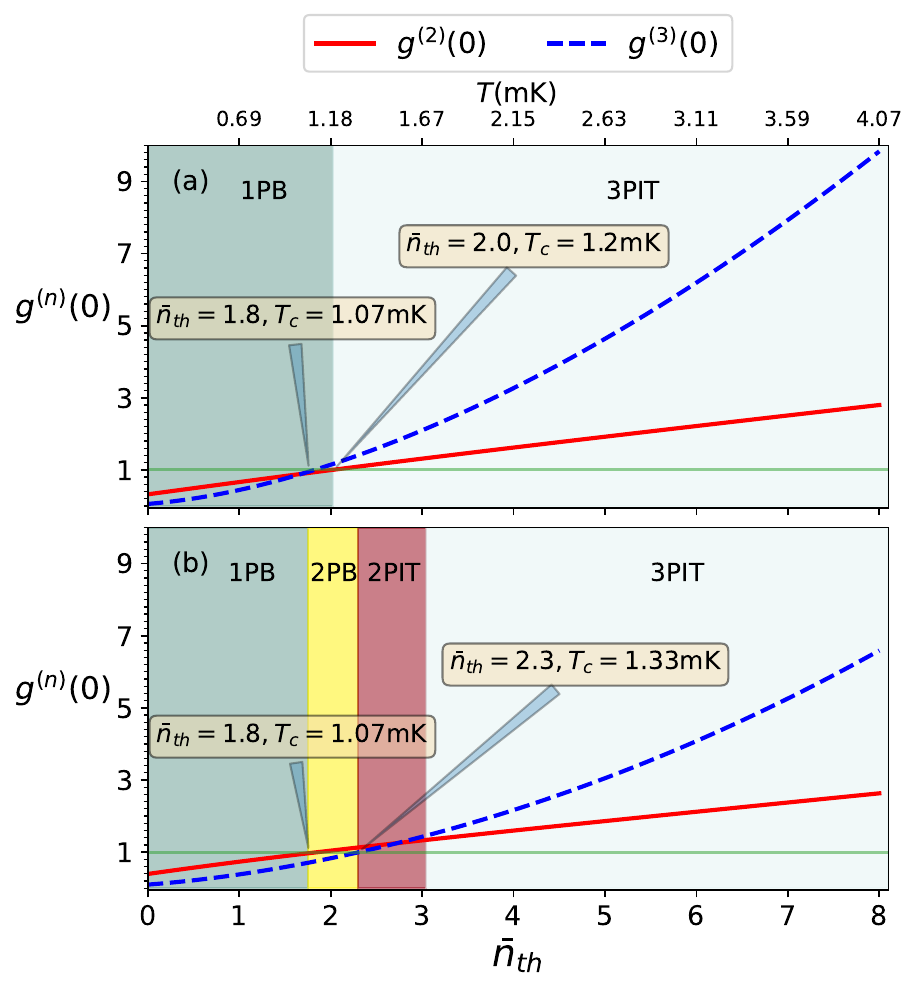}
	\caption{(Color online) The steady-state second-and third-order correlation functions versus the mechanical thermal phonon number $\bar{n}_{\rm th}$, with the same coupling strengths $g_0$,$g_{\rm CK}$, and $g_{\rm CK}'$ in (a) and (b) as those chosen in Figs.~\ref{fig5}(a) and~\ref{fig5}(b), respectively.
	The red-solid and blue-dashed lines are, respectively, referred to $g^{(2)}(0)$ and $g^{(3)}(0)$.	
	The cavity-laser detuning is zero ($\Delta_c = 0$), and
	other system parameters are the same as in Fig.~\eqref{fig4}.}
	\label{fig8}
\end{figure}

The behaviours of the steady-state second-and third-order correlation functions versus the thermal fluctuations of the phonon bath, and for different coupling regimes are presented in Fig.~\eqref{fig8}. As is seen, thermal phonons can considerably affect the correlation function $g^{(2)}(0)$ and tend to destroy 1PB. Specifically, 1PB happens below a critical temperature $T_c$  whose value depends on 
the coupling strengths $g_0, g_{\rm CK}$, and $g_{\rm CK}'$
which are in turn determined by the ratio $E_J/E_C$ as well as $\delta n_{g0}$.
As can be seen in Fig.~\ref{fig8}(a), for the coupling regime $(g_0=-9\text{MHz}, g_{\rm CK}=-0.6g_0, g_{\rm CK}'=-0.1g_0)$ corresponding to $E_J/E_C=1/20$ and $\delta n_{g0}=0.533$ the critical temperature is $T_c=1.2 \text{mK}$ (i.e., $\bar{n}_{th}=2.0$), while Fig.~\ref{fig8}(b) shows that for the coupling regime $(g_0=-10.7\text{MHz}, g_{\rm CK}=-0.9g_0, g_{\rm CK}'=-0.4g_0)$ corresponding to $E_J/E_C=1/30$ and $\delta n_{g0}=0.527$ the critical temperature is $T_c=1.07 \text{mK}$ (i.e., $\bar{n}_{th}=1.8$).
Moreover, the correlation functions $g^{(2)}(0)$ and $g^{(3)}(0)$ exhibit different behaviours with respect to temperature in the two mentioned coupling regimes. 
Figure~\ref{fig8}(a) shows that with increasing the temperature a direct transition from 1PB to 3PIT takes place, while according to Fig.~\ref{fig8}(b) this transition is mediated by the appearing the 2PB and 2PIT effects. Therefore, in the system under consideration, the quantum statistics of  photons (photon bunching and antibunching) can be controlled through adjusting the coupling regime or, equivalently, by tuning the ratio $E_J/E_C$ and $\delta n_{g0}$. 

\section{generation of the Mechanical Schrödinger cat state }\label{section:4}
In addition to the PB and PIT effects discussed so far, generation of the Schrödinger cat state is another interesting phenomenon in few-photon optomechanics. In recent years, several theoretical proposals have been made to realize Schrödinger cat states in optomechanical systems based on various mechanisms including conditional measurement on the optical field \cite{hoff2016}, photon hopping in a two-mode optomechanical system \cite{xiong2019}, utilizing the combined effects of nonlinear dynamics and dissipation \cite{tan2013}, coherent feedback in the absence of laser driving \cite{wei2021}, modulating the optomechanical coupling frequency to near resonance \cite{zheng2020}, coherently driving a quadratically coupled optomechanical cavity \cite{xie2019}, and exploiting the photon-phonon CK interaction in an undriven \cite{zou2019} as well as a strongly driven optomechanical cavity \cite{liao2020}.

In this section, we discuss the generation of the Schrödinger cat states of the mechanical-like mode in the system under consideration, benefiting from the quadratic phonon-number-dependent CK coupling $g_{\rm CK}'$, in the regime of parameters corresponding to zero optomechanical coupling ($g_0=0$). 
This is while all the above-mentioned schemes rely on the optomechanical coupling between the cavity and mechanical modes.
The scenario under consideration arises when the gate charge deviation is assumed to be equal to $\delta n_{g0}=0.5$ (see Fig.~\eqref{fig2}).
Here, we choose $E_J/E_C=1/4$, from which it follows that $g_0=0, g_{\rm CK}=-2.7 \text{MHz}$, and $g_{\rm CK}'=0.2 \text{MHz}$.
Furthermore, the conditions expressed by Eqs.~\eqref{general-condition} and~\eqref{condition5} (without considering $g_0$) are still fulfilled by these specific coupling constants.
In the absence of the driving field, the dynamical evolution of the system is governed by the unitary operator 
\begin{align}\label{Evolution}
&\hat{U}(\tau)=
\nonumber
\\
&\exp[
-i\tau(
\omega_c\hat{a}^\dagger\hat{a}+\omega_M\hat{b}^\dagger\hat{b}
+\bar{g}_{\rm CK}\hat{a}^\dagger\hat{a}\hat{b}^\dagger\hat{b}+g_{\rm CK}'\hat{a}^\dagger\hat{a}(\hat{b}^\dagger\hat{b})^2
)
].
\end{align}
We assume that the system is initially prepared in the state 
\begin{equation}\label{initial-number-state}
\ket{\psi(0)}=
\ket{n}_a \otimes \ket{\xi}_b,
\end{equation}
where $\ket{n}_a(n>0)$ and $\ket{\xi}_b= e^{-\abs{\xi}^2/2}\sum_{m=0}^{\infty}(\xi^m/\sqrt{m!})\ket{m}_b$ denote, respectively, the number state of the cavity field and the coherent state of the mechanical oscillator. Experimentally, the coherent state of the mechanical oscillator can be achieved via state transfer between an ancillary cavity mode and the mechanical oscillator, which are linearly coupled to each other \cite{verhagen2012,palomaki2013}.
In terms of the time evolution operator \eqref{Evolution}, the state of the system at time $\tau$ can be obtained as
\begin{eqnarray}\label{psi3}
\ket{\psi(\tau)}=&&\hat{U}(\tau)\ket{\psi(0)}
\nonumber
\\
&&\eta_n(\tau)e^{-\abs{\xi}^2/2}\sum_{m=0}^{\infty}\frac{\xi^{\prime m}_n(\tau)}{\sqrt{m!}}e^{-i\chi_n(\tau) m^2}
\ket{m}_b\ket{n}_a,
\end{eqnarray}
where
\begin{subequations}\label{definition}
	\begin{gather}
	\eta_n(\tau)=e^{-i\tau\omega_c n},
	\\	
	\xi^\prime_n(\tau)=\xi e^{-i\tau(\omega_M+\bar g_{\rm CK}n)},
	\\
	\chi_n(\tau)= g_{\rm CK}'\tau n.
	\label{chi_c}
	\end{gather}
\end{subequations}
At times $\tau_n^{(k)}=\frac{\pi}{k g_{\rm CK}' n}$ ($k$ is an integer) the state of the system becomes
\begin{equation}\label{state}
\ket{\psi(\tau_n^{(k)})}=\eta_n(\tau_n^{(k)})\ket{\phi}_b^{(k)}\ket{n}_a,
\end{equation}
where
\begin{equation}\label{phi}
\ket{\phi}_b^{(k)}=e^{-\abs{\xi}^2/2}\sum_{m=0}^{\infty}\frac{\xi^{\prime m}_n(\tau_n^{(k)})}{\sqrt{m!}}e^{-i\pi m^2/k}
\ket{m}_b,
\end{equation}
is the so-called Yurke-Stoler-like state \cite{yurke1986}.
Different values of the integer $k>1$ result in a variety of quantum superposition of coherent states of the mechanical-like mode.
For $k=2$, we have  
\begin{equation}\label{k2}
e^{-i\pi m^2/2}=\frac{1}{2}(1-i)+\frac{(-1)^m}{2}(1+i),
\end{equation}
and the state $\ket{\phi}_b^{(k)}$ becomes
\begin{equation}\label{state2}
\ket{\phi}_b^{(2)} = \frac{1-i}{2}\ket{\xi_n'(\tau_n^{(2)})}_b+\frac{1+i}{2}\ket{-\xi_n'(\tau_n^{(2)})}_b,
\end{equation}
which is a two-component Schrödinger cat state of the mechanical oscillator. When $k=3$, we have 
\begin{equation}\label{k3}
e^{-i\pi m^2/3}=a_1e^{-im\pi/3}+a_2e^{im\pi/3}+a_3(-1)^m,
\end{equation}
where $a_1=a_2=\frac{1}{3}(1+e^{-i\pi/3}) $ and $a_3=\frac{1}{3}(1-2e^{-i\pi/3})$.
In this case, a three-component (unnormalized) mechanical cat state can be obtained 
\begin{equation}\label{state3}
\ket{\phi}_b^{(3)} = a_1\ket{\xi_n'(\tau_n^{(3)})e^{-i\pi/3}}_b+
a_2\ket{\xi_n'(\tau_n^{(3)})e^{i\pi/3}}_b+
a_3\ket{-\xi_n'(\tau_n^{(3)})}_b,
\end{equation}
By taking $k=4$, we have
\begin{equation}\label{k4}
e^{-i\pi m^2/4}=\frac{1}{2}e^{-i\pi/4}[1-(-1)^m]+\frac{(-1)^m}{2}i^m[1+(-1)^m],
\end{equation}
and we obtain a cat state with four superposition components as
\begin{align}\label{state4}
\ket{\phi}_b^{(4)} &= \frac{1}{2}e^{-i\pi/4}
[\ket{\xi_n'(\tau_n^{(4)})}_b-\ket{-\xi_n'(\tau_n^{(4)})}_b]
\nonumber
\\
&+
\frac{1}{2}
[\ket{-i\xi_n'(\tau_n^{(4)})}_b+\ket{i\xi_n'(\tau_n^{(4)})}_b].
\end{align}
The above results can be generalized to conclude that at time $\tau_n^{(k)}$ the $k$-component mechanical Schrödinger cat state $\ket{\phi}_b^{(k)}$ is generated. 

The signature of nonclassical characteristic of the generated mechanical Schrödinger cat states $\ket{\phi}_b^{(k)}$ can be revealed by calculating their corresponding Wigner functions defined as \cite{schleich} 
\begin{equation}\label{Wigner_dis}
W(x,p)=\frac{1}{2\pi}\int_{-\infty}^\infty dy \bra{x+\frac{y}{2}}\hat{\rho}_b^{(k)}\ket{x-\frac{y}{2}}\exp(-ipy),
\end{equation}
where $\hat{\rho}_b^{(k)}=\ket{\phi}_b^{(k)}  \prescript {(k)}{b}{\bra{\phi}}$, and $x$ and $p$ are the position and momentum variables in the phase space, respectively.

\begin{figure*}[t]
	\centering
	\includegraphics[width=17.2cm]{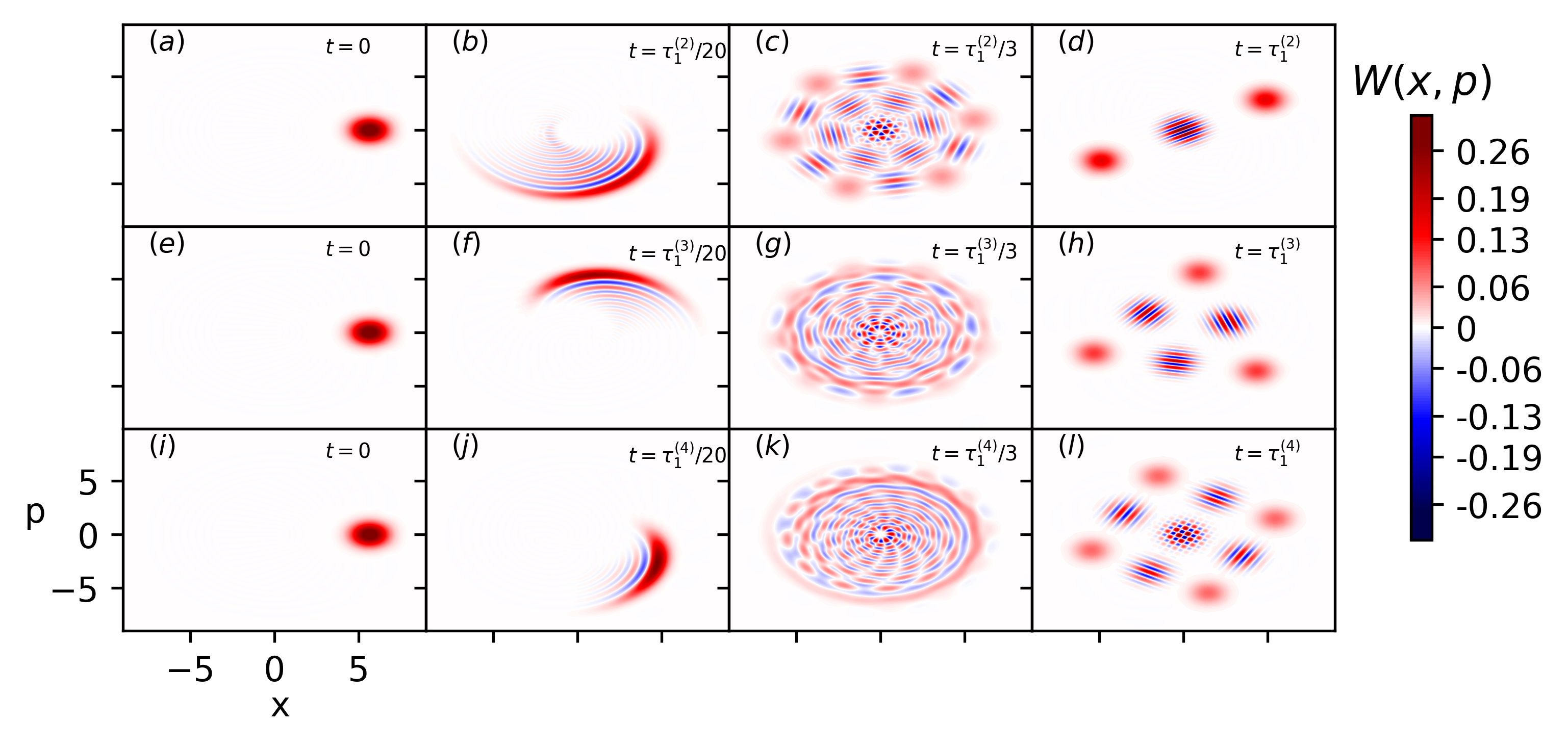}
	\caption{ (Color online) 
	Numerical simulations of the time evolution of the Wigner functions $W(x,p)$ of the generated states $\ket{\phi}_b^{(k)}$ versus dimensionless canonical quadratures $x$ and $p$ for $k=2,3,4$, and $n=1$ over the time range $t=0$ to $\tau_1^{(k)}$. The relevant parameters are set as $\omega_c/2\pi=5\text{GHz}$, $\omega_M/2\pi=10\text{MHz}$, $g_0=0$, $g_{\rm CK}=-2.7\text{MHz}$, $g_{\rm CK}'=0.2\text{MHz}$, and $\xi = 4$.
	}
	\label{fig9}
\end{figure*}

\begin{figure}[t]
	\centering
	\includegraphics[width=8.6cm]{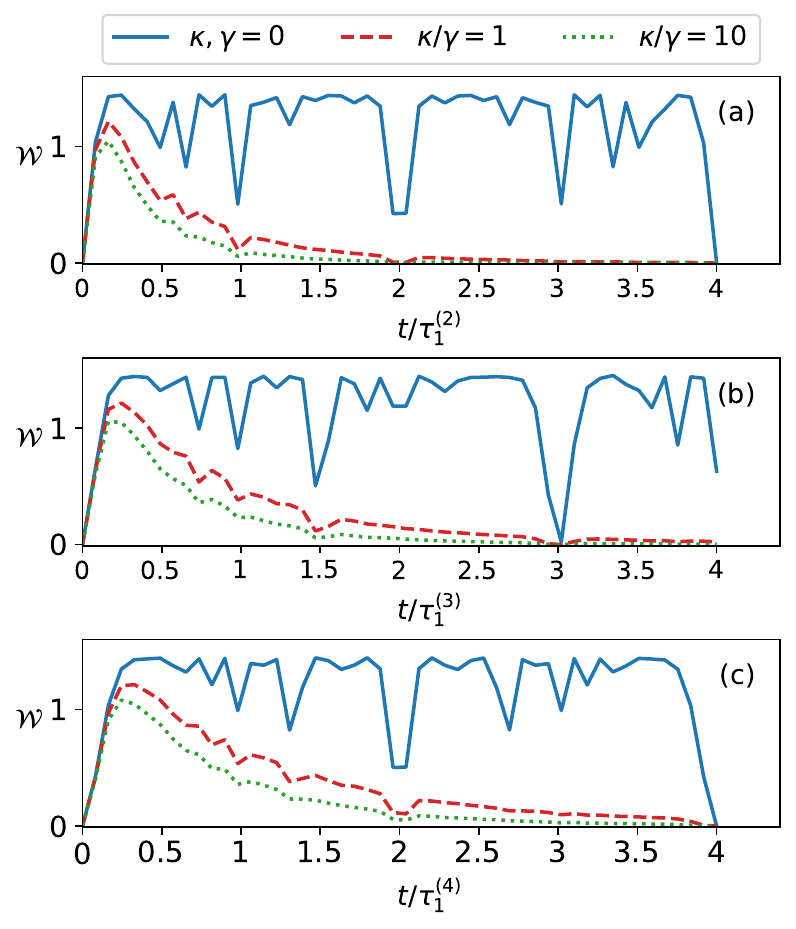}
	\caption{(Color online) Time evolution of the Wigner negativity for (a) $k=2$, (b) $k=3$, and (c) $k=4$.  
	Here, $\gamma=10$ kHz, $\xi=4$ and $\bar{n}_{\rm th} = 0$, and other system parameters are the same as in Fig.~\eqref{fig9}.}
	\label{fig10}
\end{figure}

In Fig.~\eqref{fig9}, by using QUTIP, we have simulated the time evolution of the Wigner functions for the states $\ket{\phi}^{(k)}_b$ with $k=2,3,4$ when $n=1$. As can be seen, the state of the mechanical oscillator evolves from the initial coherent state $\ket{\xi}_b$ to a $k$-component Schrödinger cat-like state at time $\tau_1^{(k)}$ (see Figs.~\ref{fig9}(d), (h), and (l)).
Therefore, the numerical and analytical results are perfectly identical to each other.
It should be noted that the second-order dispersive shift $g_{\rm CK}'$ realized by our scheme is responsible to create the macroscopically distinct superposition states in the mechanical-like mode.
This can be easily seen from Eqs.~\eqref{psi3} and~\eqref{chi_c} as the mechanical oscillator would be remained in its initial coherent state (up to a phase factor) if $g_{\rm CK}'$ was not taken into account (i.e., in case of neglecting the quadratic-quartic interaction between the cavity and mechanical modes, see Eq.~\eqref{gck_def}).

To investigate the influence of the system dissipation on the Wigner functions of the generated mechanical Schrödinger cat-like states, we use the Wigner negativity $\mathcal{W}$ which is defined as \cite{kenfack2004}
\begin{equation}\label{Wigner-negativity}
\mathcal{W}=\iint[\abs{W(x,p)}-W(x,p)]dxdp.
\end{equation} 
This is the volume of the negative part of the Wigner distribution, such that $\mathcal{W}\ge0$, and is often considered as a quantumness (nonclassicality) measure. By definition the quantity $\mathcal{W}$ is equal to zero for classical states, while $\mathcal{W}>0$ corresponds to nonclassical states.
Figure~\eqref{fig10} shows numerically calculated time evolution of the Wigner negativity for two different rates of cavity dissipation, namely $ \kappa=10$ kHz (red-dashed lines), and $\kappa=100$kHz (green-dotted lines), and also for the ideal (free-dissipation) case $\kappa, \gamma =0$ (blue-solid lines).
The Wigner negativity in the absence of dissipation oscillates during the time $t/\tau_1^{(k)}$ for $k=2,3,4$. 
However, in the presence of the dissipation, as time goes on, the effects of decoherence become increasingly prominent, leading to the eventual disappearance of the Wigner negativity. 
Moreover, the nonclassicality of the generated mechanical states is suppressed more rapidly as $\kappa/\gamma$ increases. Here, we should emphasize that the negativity of a Wigner distribution is not sufficient to imply the existence of a Schrödinger cat state; it merely indicates the nonclassicality of the state.
Note that $\tau_1^{(2)}=7.8 \mu s$, $\tau_1^{(3)}=5.2 \mu s$, and $\tau_1^{(4)}=3.9 \mu s$ are close to the cavity decay time $(\kappa^{-1}\approx 10 \mu s)$. This provides a suitable experimental situation to monitor the nonclassical state.
Most interestingly, as can be seen from Fig.~\ref{fig10}, the generated mechanical cat states are robust against dissipation (see the red dashed and green dotted curves as $t/\tau_1^{(k)}=1 (k=2,3,4)$).
It should be noted that for the steady-state mechanical cat state to emerge one needs to cool down the mechanics to near its ground state where $ n_{th}\simeq 0 $, since the presence of thermal noises immediately destroys the nonclassical behavior of the system.

\section{microwave-mechanics entanglement}\label{section:5}
In this section, we would like to investigate the entanglement between the microwave and mechanical modes in the presence of ordinary and generalized CK nonlinearities. 
To this aim, we should solve the steady state of the linearized quantum Langevin equations of motion using Hamiltonian~\eqref{H_total} by  considering a strong cavity driving laser term as \cite{vitalli-entanglement} (see Appendix \ref{appendix3})
\begin{subequations}\label{Langevin}
\begin{eqnarray}
&& \delta\dot{\hat{a}}=-(i\Delta_{\rm eff}+\kappa)\delta\hat{a}+i\frac{g_{\rm eff}}{2}(\delta\hat{b}^\dagger+\delta\hat{b})+\sqrt{2\kappa}\hat{a}_{\rm in},\label{Lan1} \\
&& \delta\dot{\hat{b}}=-(i\omega_{\rm eff}+\gamma)\delta\hat{b}+i\frac{g_{\rm eff}}{2}(\delta\hat{a}^\dagger+\delta\hat{a})+\sqrt{2\gamma}\hat{b}_{\rm in}.  \label{Lan2}
\end{eqnarray}
\end{subequations}
Here $\Delta_{\rm eff}=\Delta_c+g_0(\beta+\beta^*)+\tilde{g}_{\rm CK}\abs{\beta}^2+g_{\rm CK}'\abs{\beta}^4$ is the effective cavity detuning and $\omega_{\rm eff}=\omega_M+\tilde{g}_{\rm CK}\abs{\alpha}^2+2g_{\rm CK}'\abs{\alpha}^2\abs{\beta}^2$ is the effective frequency of the mechanical mode.
Also $g_{\rm eff}=-2g\abs{\alpha}$ is the effective optomechanical coupling strength with $g=g_0+\tilde{g}_{\rm CK}\beta+2g_{\rm CK}'\beta^3$, and $\tilde{g}_{\rm CK}=g_{\rm CK}+2g_{\rm CK}'$.
The equations of motion~\eqref{Langevin} show that in the
linearized regime, the dynamics of the system in the Schrödinger picture is governed by the effective Hamiltonian
\begin{gather}
\hat{H}_{\rm eff}^{(2)}=\Delta_{\rm eff} \delta\hat{a}^\dagger\delta\hat{a}+\omega_{\rm eff}\delta\hat{b}^\dagger\delta\hat{b}-\frac{g_{\rm eff}}{2}(\delta\hat{a}^\dagger+\delta\hat{a})(\delta\hat{b}^\dagger+\delta\hat{b})
\nonumber
\\
-i\kappa\delta\hat{a}^\dagger\delta\hat{a}-i\gamma\delta\hat{b}^\dagger\delta\hat{b}.
\label{H-linear-effective}
\end{gather}

If we define the quadratures $\delta \hat{Q}=(\delta\hat{b}^\dagger+\delta\hat{b})/\sqrt{2}$, $\delta \hat{P}=(\delta\hat{b}-\delta\hat{b}^\dagger)/i\sqrt{2}$ for the resonator, and $\delta \hat{X}=(\delta\hat{a}^\dagger+\delta\hat{a})/\sqrt{2}$, $\delta \hat{Y}=(\delta\hat{a}-\delta\hat{a}^\dagger)/i\sqrt{2}$ for the cavity mode, Eqs.~\eqref{Langevin} can be expressed in the compact matrix form
\begin{equation}\label{compact}
\dot{\hat{u}}(t)=A\hat{u}(t)+\hat{N}(t),
\end{equation}
where $\hat{u}^T(t)=(\delta \hat{Q}(t),\delta \hat{P}(t),\delta \hat{X}(t),\delta \hat{Y}(t))$ is the vector of continuous variables (CV) fluctuation operators, $\hat{N}^T(t)=(\sqrt{2\gamma}\hat{Q}_{\rm in}(t),\sqrt{2\gamma}\hat{P}_{\rm in}(t),\sqrt{2\kappa}\hat{X}_{\rm in}(t),\sqrt{2\kappa}\hat{Y}_{\rm in}(t))$  is the vector of noises, and the drift matrix is
\begin{equation}\label{matrixA}
A=
\begin{pmatrix}
-\gamma & \omega_{\rm eff} & 0 & 0
\\
-\omega_{\rm eff} & -\gamma & g_{\rm eff} &0
\\
0 & 0 & -\kappa & \Delta_{\rm eff}
\\
g_{\rm eff} & 0 & -\Delta_{\rm eff} & -\kappa
\end{pmatrix}.
\end{equation}

Since the dynamics of the system is linearized and the quantum noises are zero-mean quantum Gaussian noises, the steady state of the system is a Gaussian bipartite state \cite{vitalli-entanglement}. Therefore the steady state of the system is characterized by its $4\times4$ correlation matrix \cite{weedbrook2012}
\begin{equation}\label{V}
V_{ij}=(\expval{\hat{u}_i(\infty)\hat{u}_j(\infty)+\hat{u}_j(\infty)\hat{u}_i(\infty)})/2,
\end{equation}
where $\hat{u}^T(\infty)=(\delta \hat{Q}(\infty),\delta \hat{P}(\infty),\delta \hat{X}(\infty),\delta \hat{Y}(\infty))$ is the vector of CV fluctuation operators at the steady state $(t\rightarrow\infty)$.

According to the Routh-Hurwitz criterion \cite{DeJesus1987}, the system is stable only if the real part of all the eigenvalues of matrix A is negative.
In Appendix~\ref{appendix4}, we examine the Routh-Hurwitz criterion in the system.
Therefore, when the stability conditions are fulfilled, the steady state of the correlation matrix can be obtained by solving the Lyapunov equation 
\begin{equation}\label{Lyapunov}
AV+VA^T=-D,
\end{equation}
where $\hat{D}^T=\text{diag}[\gamma(2\bar{n}_{\rm th}+1),\gamma(2\bar{n}_{\rm th}+1),\kappa,\kappa]$ is the diagonal diffusion matrix.
However, the analytical solution to Eq.~\eqref{Lyapunov} for the correlation matrix $V$ is very cumbersome, so we mainly adopt the numerical simulations. The microwave-mechanical bipartite entanglement can be quantified by using the logarithmic negativity $E_N$. For Gaussian states it reads\cite{adesso2004}
\begin{equation}\label{negativity}
E_N=\max[0,-\ln(2\nu^-)].
\end{equation}
where $\nu^-=2^{-1/2}\{\Sigma(V)-[\Sigma(V)^2-4\det V]^{1/2}\}^{1/2}$ is the lowest
symplectic eigenvalue of the partial transpose of the correlation matrix with $\Sigma(V)=\det V_C+ \det V_M - 2\det V_{CM}$, and we have used the $2\times2$ block form of
\begin{equation}\label{V2}
V=\begin{pmatrix}
V_M & V_{CM} \\ V_{CM}^T & V_C 
\end{pmatrix}.
\end{equation}
Here $V_M$ and $V_C$ are, respectively, correspond to the mechanical and the cavity modes, and $V_{CM}$ is related to the optomechanical correlation.
According to Eq.~\eqref{negativity}, when $\nu^-<1/2$ the state in question will exhibit entanglement.

\begin{figure}[t]
	\centering
	\includegraphics[width=8.6cm]{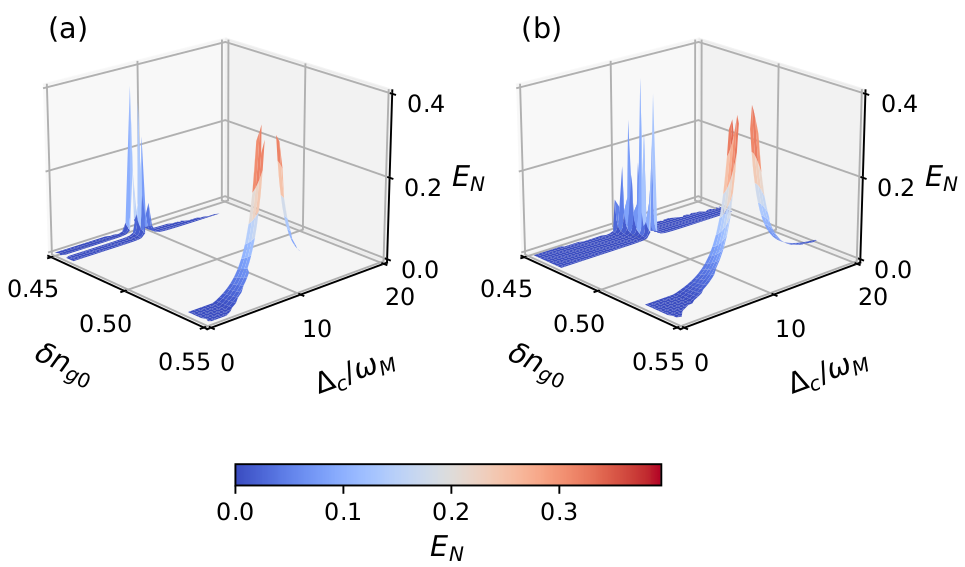}
	\caption{
	(Color online) Logarithmic negativity $E_N$ as a function of cavity detuning $\Delta_c/\omega_M$ and gate charge deviation $\delta n_{g0}$ for (a) $E_J/E_C=1/4$, and (b) $E_J/E_C=1/5$.	
	The other parameters are $\omega_c/2\pi=10$ GHz, $\omega_M/2\pi=50$ MHz, $\mathcal{P}=-50$ dBm, $\kappa= 1$ MHz, $\gamma= 500$ kHz, and $\bar{n}_{\rm th}= 0.5$.
	}
	\label{fig11}
\end{figure}

The logarithmic negativity $ E_N $, plotted as a function of cavity detuning and gate charge deviation $\delta n_{g0}$, is demonstrated in Fig.~\eqref{fig11} for two distinct coupling regimes: $(a)$ $E_J/E_C=1/4$, and $(b)$ $E_J/E_C=1/5$.
Here, we interestingly investigate the entanglement in the red-detuned sideband $(\Delta_c>0)$ notably in the dispersive red-detuning corresponding to the large detuning. 
However, the entanglement is still fragile to the thermal noise, which means that the mechanical oscillator must be precooled down to near its ground state $(\bar{n}_{\rm th}=0.5)$.
Figure~\eqref{fig11} shows that for the selected initial parameters, the higher values of $\delta n_{g0}$ lead to the enhancement of microwave-mechanics entanglement. 

As is seen, Fig.~\eqref{fig11} does not include results for the logarithmic negativity in the middle range of $\delta n_{g0}$ values. This is because of certain restrictions imposed on the system parameters. The first restriction comes from conditions \eqref{condition5} and \eqref{general-condition} which were also taken into account in Fig.~\eqref{fig4}. The second restriction concerns the linearization conditions, $\abs{\alpha},\abs{\beta}\gg1$, which are realized when the driving power is large enough.
The third and final restriction is due to the stability condition for the system, which is justified according to the Routh-Hurwitz criterion ( see Appendix~\ref{appendix4}).

\begin{figure}[t]
	\centering
	\includegraphics[width=8.6cm]{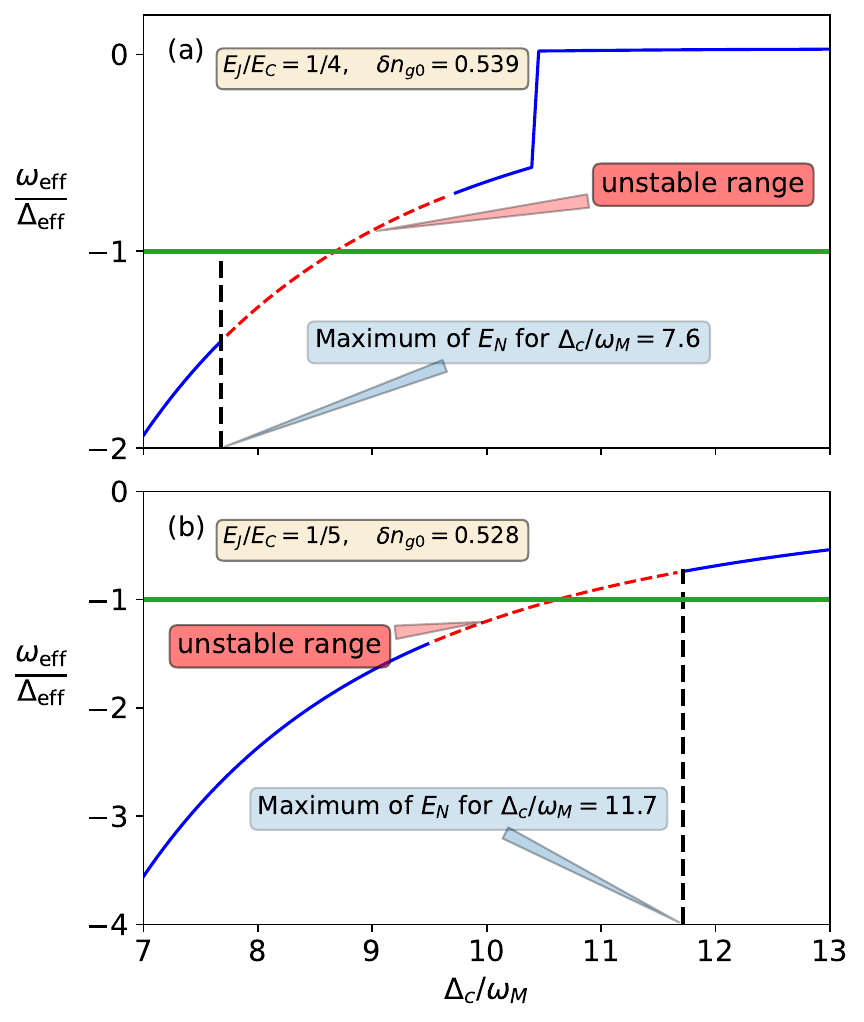}
	\caption{(Color online) The scaled effective frequency $\omega_{\rm eff}/\Delta_{\rm eff}$ as a function of cavity detuning $\Delta_c/\omega_M$ for (a) $E_J/E_C=1/4, \delta n_{g0}=0.539$ and (b) $E_J/E_C=1/5, \delta n_{g0}=0.528$.
	The green lines show $\omega_{\rm eff}/\Delta_{\rm eff}=-1$. 
	The dashed red lines denote unstable range according to the Routh-Hurwitz condition.
	The black solid lines show the maximum values of logarithmic negativity.
	The other system parameters are the same as in Fig.~\eqref{fig11}.}
	\label{fig12}
\end{figure}

Moreover, the value of logarithmic negativity $E_N$ depends on the system parameters $\Delta_{\rm eff},\omega_{\rm eff},g_{\rm eff}$ in Eq.~\eqref{H-linear-effective}. 
The investigation of the entanglement is complicated due to the dependence of these values on the coupling strengths $(g_0,g_{\rm CK},g_{\rm CK}')$ in Eq.~\eqref{H_total} and can be controlled by detuning parameter $\Delta_c$.
To clarify this matter, firstly in Fig.~\eqref{fig12}, the ratio of $\omega_{\rm eff}/\Delta_{\rm eff}$ is plotted versus $\Delta_c/\omega_{m}$.
Panels (a) and (b) in this figure are, respectively, for $\delta n_{g0}=0.539$ and $\delta n_{g0}=0.528$.

The red dashed lines in both panels indicate the region where the system is unstable according to the Routh-Hurwitz criterion and the vertical black dashed lines correspond to the detunings at which the maximum of logarithmic negativity occurs. In addition, $\Delta_{\rm eff} = -\omega_{\rm eff}$ is shown with a green horizontal solid line. 
The maximum values of the logarithmic negativity are obtained for $\Delta_c/\omega_{M}=7.6$ in Fig.~\ref{fig12}(a) and for $\Delta_c/\omega_{M}=11.7$ in Fig.~\ref{fig12}(b) which both of them are adjacent to $\Delta_{\rm eff} = -\omega_{\rm eff}$. These results are consistent with those expected for linearized optomechanical systems \cite{aspelmeyer2014,milburnbook}. 

\begin{figure}[t]
	\centering
	\includegraphics[width=8.6cm]{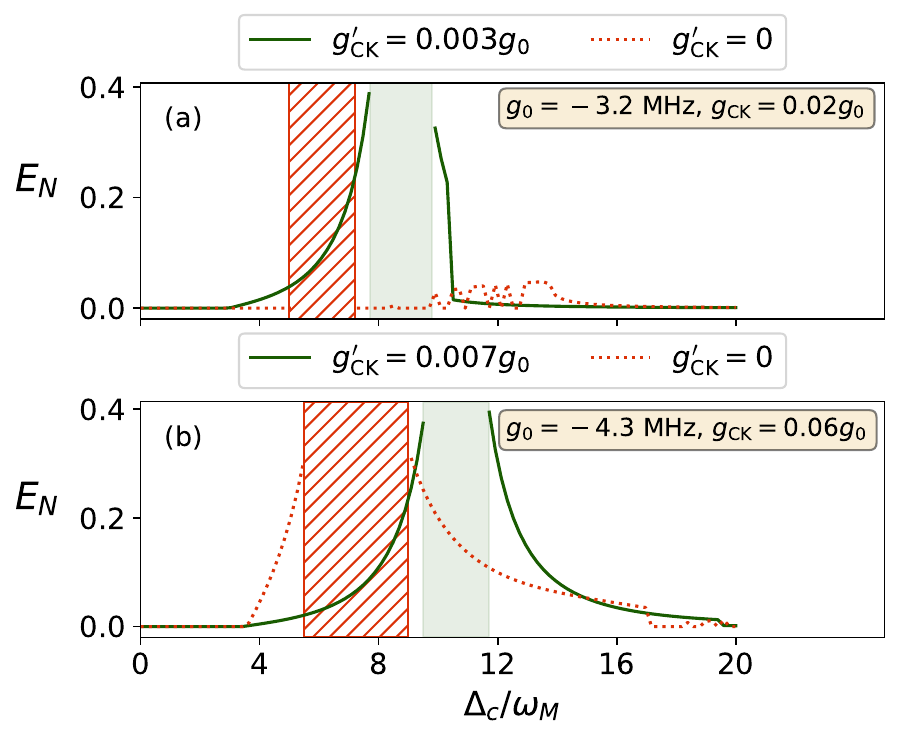}
	\caption{
	(Color online) The logarithmic negativity as a function of cavity detuning $\Delta_c/\omega_M$ for the parameters regimes considered in Fig.~\eqref{fig12}, and for both cases of with and without the generalized CK term.
    The green and red hatched boxes, respectively, show unstable regions for the green-solid and red-dotted lines according to the Routh-Hurwitz condition.}
	\label{fig13}
\end{figure}

As mentioned above, the logarithmic negativity depends on the coupling strengths. 
Therefore, we are interested to know how the generalized CK Hamiltonian term affects the entanglement in the system. 
We thus plot in Figs.~\ref{fig13} (a) and (b) the logarithmic negativity versus the cavity detuning $\Delta_c/\omega_M$ for the parameter regimes considered, respectively, in Figs.~\ref{fig12} (a) and (b), and for both cases of with and without the generalized CK term. As can be seen from Fig.~\ref{fig13}(a), for $g_{\rm CK}^\prime\ne0$ the photon-phonon entanglement is significantly enhanced.
In Fig.~\ref{fig13}(b), while there are slight variations in the entanglement values, the primary distinction lies in the shift of the entanglement degree relative to the detuning parameter $\Delta_c/\omega_{M}$.

\section{Experimental Discussion}\label{section:6}
The experimental realization of the tripartite microwave optomechanical system discussed in this paper has been reported in Ref.~\cite{pirkkalainen2015}.
The cavity and mechanical frequencies, determined by Eqs.~\eqref{A10a}~and~\eqref{A10b}, have been here chosen to be $\omega_c/2\pi=5-10\text{GHz}$ and $\omega_M/2\pi=10-50\text{MHz}$ which are experimentally realistic. The cavity and mechanical decay rates depend on the gate charge \cite{pirkkalainen2015}, and we have considered them to be in the ranges of $10\text{kHz}-1\text{MHz}$ and $10\text{kHz}-500\text{kHz}$, respectively, which are accessible with the current experimental situations. For the numerical simulations of the photon blockade and the microwave-mechanic entanglement the values of the driving laser power have been, respectively, set to $\mathcal{P}=-155$ and $\mathcal{P}=-50$ dBm, which are in the range used in some recent experiments \cite{pirkkalainen2015,brookes, astafiev2007}. 
In our calculations, the gate voltage is changeable in the range of $V_g =(1-10)$V, the capacitances are $C=50 $fF, $C_{g_0}=(0.1-1)$fF, $C_1+C_2=(0.2-2)$fF, the Josephson energy $E_J/\hbar=10$ GHz, the inductance $L=(5-15)$nH, and $\frac{4 Z_0 e^2}{h}\approx10^{-2}$ ($ Z_0=\sqrt{L/C} $) \cite{pirkkalainen2015,heikkila2014}. 
As the final point, based on the experimental parameters \cite{pirkkalainen2015,heikkila2014} the accessibility of the precision of the deviation gate charge is considered $ \Delta (\delta n_{g_0}) < 10^{-3}$ that implies high-stable and precise gate voltage and the gate capacitance with relative errors $ \Delta Y/Y \lesssim 10^{-8}-10^{-9} $ ($ Y=V_g, C_{g_0} $).

\section{Conclusion and outlook}\label{section:7}
In the present work, we have investigated the few-photon optomechanical effects, including photon blockade and mechanical Schrödinger cat state generation, as well as photon-phonon entanglement in a generalized microwave optomechanical system where in addition to the strong optomechanical and CK photon–phonon couplings, a higher-order nonlinear CK coupling is present. We found that in the presence of the higher-order CK coupling the degrees of photon bunching and antibunching can be effectively enhanced and controlled by adjusting the system parameters. Interestingly, we have also shown that in the regime of zero optomechanical coupling the generalized CK nonlinearity can result in the existence of multi-component mechanical superposition states which are robust against the system dissipation. Moreover, our results reveal that the generalized CK nonlinearity can enhance substantially the photon-phonon entanglement in the regime of large-red detuning.

As an outlook for further study, the system can be investigated for other nonclassical properties such as phonon or photon-photon blockade, quadrature squeezing, mechanical cooling, and normal mode splitting. Furthermore, the chaotic dynamics of the system might be an interesting subject for future investigation.

\section*{ACKNOWLEDGMENTS and Contributions}
A.M. is grateful the Office of the Vice President for Research of the University of Isfahan and ICQT. 
A.M. would also thanks Professor David Vitali, whose work on CK optomechanics provided motivation me to extend and generalize it.
H.S. wishes to thank Professor Rasoul Roknizadeh, of the CQST at the University of Isfahan, for support.

A.M. proposed and developed the primary idea of formulating the generalized CK nonlinearity and its corresponding experimental scheme as well as its effect on the 2PB and photon-phonon entanglement. MHN proposed the PIT and generation of the multicomponent mechanical cat state in this system.
All analytical and numerical calculations were performed by H.S., and rechecked by A.M. and M.H.N. H.S. developed the idea of characterizing the nonclassical state by using the Wigner negativity. 
All authors contributed to the discussion of the results and the writing of the manuscript.

\appendix

\section{Derivation of Hamiltonian \eqref{H1}}\label{appendix1}
We start with the Hamiltonian~(6) in Ref.~\cite{heikkila2014} describing a single-Cooper-pair transistor (SCPT) which consist of a microwave cavity and a mechanical resonator coupled to a common qubit:
\begin{gather}
\hat{H}_t=  \hat{H}_{\rm SCPT} + \hbar \omega_c^0 \hat{c}^\dagger\hat{c} + \hbar \omega_M^0\hat{d}^\dagger\hat{d}
+ g_m \hat{\sigma_3} \hat{x}_m 
\nonumber
\\
+(g_{q1} \hat{\sigma}_1 + g_{q2} \hat{\sigma}_2)\hat{x}_c^2
\nonumber
\\
+(g_{c1} \hat{\sigma}_1 + g_{c2} \hat{\sigma}_2)\hat{x}_c,
\label{H-appendix1}
\end{gather}
where
\begin{subequations}
\begin{gather}
\hat{H}_{\rm SCPT} = \frac{B_1}{2} \hat{\sigma}_1 + \frac{B_2}{2} \hat{\sigma}_2 +\frac{B_3}{2} \hat{\sigma}_3,
\label{A2a}
\\
g_m =-\frac{4E_Cx_{xp}(\partial_xC_g)V_g}{2e},
\\
g_{q1}=\frac{e^2Z_0}{8\hbar}(E_{J2}+E_{J2})\cos(\phi_a/2),
\label{A2c}
\\
g_{q2}=\frac{e^2Z_0}{8\hbar}(E_{J2}-E_{J1})\sin(\phi_a/2),
\label{A2d}
\\
g_{c1}=\sqrt{\frac{e^2Z_0}{8\hbar}}(E_{J2}+E_{J2})\sin(\phi_a/2),
\label{A2e}
\\
g_{c2}=\sqrt{\frac{e^2Z_0}{8\hbar}}(E_{J2}-E_{J2})\cos(\phi_a/2).
\label{A2f}
\end{gather}
\label{H-SCPT}
\end{subequations}
Here, the quantities $\omega_c^0$ and $\omega_M^0$ denote, respectively, the natural frequencies of the cavity and the mechanical modes, with respective annihilation (creation) operators $\hat{c} (\hat{c}^\dagger)$ and $\hat{d} (\hat{d}^\dagger)$ and the position operators $\hat{x}_c=\hat{c}^\dagger+\hat{c}$ and $\hat{x}_d=\hat{d}^\dagger+\hat{d}$.
$\hat{\sigma}_{1}$, $\hat{\sigma}_{2}$, and $\hat{\sigma}_{3}$ are Pauli matrices for the qubit, and $E_{Ji} (i=1, 2)$ are the Josephson energies. The effective magnetic fields $B_1$, $B_2$, and $B_3$ are given by
\begin{subequations}
\begin{eqnarray}
&& B_1=-(E_{J1}+E_{J2})\cos(\phi_a/2), \label{B1} \\
&& B_2=(E_{J1}-E_{J2})\sin(\phi_a/2), \label{B2} \\
&& B_3=4E_C(1-2\delta n_{g0}), \label{B3} 
\end{eqnarray}
\end{subequations}
where $\phi_a$ is the average phase difference of the superconducting order parameters across the junction. 
In addition, $g_m, g_{q1,2}$, and $g_{c1,2}$ describe the coupling strengths between the qubit and resonators.
Finally, $e$ is the electric charge unit, and $Z_0=\sqrt{L/C}$ with $L$ and $C$ being, respectively, the geometric inductance and capacitance (see Fig.~\eqref{fig1}).

Similar to Ref.~\cite{heikkila2014}, for the sake of simplicity, we consider the case of symmetric junctions, $E_{J1}=E_{J2}=E_{J}$, in which case according to Eqs.~\eqref{A2d},~\eqref{A2f}, and~\eqref{B2}, the coupling strengths $g_{q2}$ and $g_{c2}$ as well as the magnetic field $B_2$ are all vanishing. Moreover, we set $\phi_a=0$ so that Eq.~\eqref{A2e} gives $g_{c1}=0$.  In this way the Hamiltonian Eq.~\eqref{H-appendix1} takes the form
\begin{gather}
\hat{H}_t=  \hbar \omega_c^0 \hat{c}^\dagger\hat{c} + \hbar \omega_M^0\hat{d}^\dagger\hat{d}+\frac{B_1}{2} \hat{\sigma}_1 + \frac{B_3}{2} \hat{\sigma}_3
\nonumber
\\
+ g_m \hat{\sigma_3} \hat{x}_m +
g_{q} \hat{\sigma_1}\hat{x}_c^2,
\label{H-appendix}
\end{gather}
in which $g_{q}\equiv g_{q1}$.

\section{Derivation of Hamiltonian \eqref{H_total}}\label{appendix2-2}
As mentioned in Sec.~\ref{section:2}, in the dispersive limit, $\hbar\omega_{c,M}^{(0)}\ll \abs{B}=\sqrt{B_1^2+B_3^2}$, and when all couplings are small enough one needs to diagonalize only the interaction part of the Hamiltonian \eqref{H1} in the qubit basis. In this basis, we can write
\begin{equation}\label{matrix}
\hat{H}_t=
\begin{pmatrix}
\frac{B_3}{2}+ g_m\hat{x}_m  & \frac{B_1}{2} + g_{q}\hat{x}_c^2 
\\
\\
\frac{B_1}{2} + g_{q}\hat{x}_c^2  &\quad -\frac{B_3}{2}- g_m\hat{x}_m
\end{pmatrix}.
\end{equation}
Assuming the qubit to be in its ground state, we can use the replacement $\hat{\sigma}_3\rightarrow -1$, and we find the corresponding eigenvalue of $\hat{H}_t$ as
\begin{gather}\label{eigenvalue}
\lambda = -\frac{B}{2}\big(
1+ x
\big)^{1/2},
\end{gather}
where
\begin{gather}\label{x}
x =\frac{4}{B^2}\Big(
B_1g_q\hat{x}_c^2 + B_3g_m\hat{x}_m + g_m^2\hat{x}_m^2 +g_q^2\hat{x}_c^4 
\Big). 
\end{gather}
Since $-1<x\leq1$, with the help of Binomial expansion we have
\begin{gather}
\big(1+ x\big)^{1/2}
\approx
1+\frac{1}{2}x- \frac{1}{8}x^2 + \frac{3}{48} x^3 - \frac{15}{384}x^4+\frac{21}{768}x^5.
\end{gather}
Using the multinomial theorem \cite{multinomial} together with Eqs.~\eqref{eigenvalue} and~\eqref{x} we obtain
\begin{align}
\hat{H}_t=&\hbar \omega_c^0 \hat{c}^\dagger\hat{c} + \hbar \omega_M^0\hat{d}^\dagger\hat{d}+\alpha_m\hat{x}_m
+\hbar g_{Sc} \hat{x}_c^2+\hbar g_{Sm}\hat{x}_m^2
\nonumber
\\
&+\hbar g_{\rm rp}\hat{x}_c^2\hat{x}_m+\hbar g_{ck}^0\hat{x}_c^2\hat{x}_m^2+\hbar g_{\rm cub}^0\hat{x}_c^2\hat{x}_m^3+\hbar g_{\rm qurtic}^0\hat{x}_c^2\hat{x}_m^4
\nonumber
\\
&+\hbar G_{1}^0\hat{x}_c^4\hat{x}_m+\hbar G_{2}^0\hat{x}_c^4\hat{x}_m^2+\hbar G_{3}^0\hat{x}_c^4\hat{x}_m^3+\hbar G_{4}^0\hat{x}_c^4\hat{x}_m^4,
\label{A6}
\end{align}
where
\begin{subequations}
\begin{gather}
\alpha_m = -\frac{B_3g_m}{B},
\\
\hbar g_{\rm Sc}= - \frac{g_q B_1}{B},
\\
\hbar g_{\rm Sm}= - \frac{B_1^2g_m^2}{B^3},
\\
\hbar g_{\rm rp}= 2 \frac{B_1 B_3 g_m g_q}{B^3},
\\
\hbar g^0_{\rm CK}= 2 \frac{B_1 g_m^2 g_q}{B^5}(B^2-3B_3^2),
\\
\hbar g^0_{\rm cub}=4 \frac{B_1 B_3 g_m^3 g_q }{B^7}(5 B_3^2-3B^2),
\\
\hbar g^0_{\rm quartic}= 2\frac{B_1g_qg_m^4}{B^9}(-3B^4+30(BB_3)^2-35B_3^4),
\label{A10g_duf}
\\
\hbar G_{1}^0=2 \frac{B_3 g_m g_q^2}{B^5}(B^2-3B_3^2),
\\
\hbar G_{2}^0=\frac{2g_m^2g_q^2}{B^7}(15B_1^2B_3^2-2B^4),
\label{A10G2}
\\
\hbar G_{3}^0=\frac{4g_m^2g_q^2B_3}{B^9}(5B^2B_3^2+15B_1^2B^2-3B^4-35B_3^2B_1^2),
\\
\hbar G_{4}^0=\frac{g_m^4g_q^2}{B^9}(60B^2B_3^2+30B^2B_1^2-6B^4-70B_3^4-420B_3^2B_1^2).
\end{gather}
\label{coefficient}
\end{subequations}
Note that in Eq.~\eqref{A6} the linear term with respect to $\hat{x}_c$ is absent because of the assumption $\phi_a=0$.

In the derivation of Hamiltonian \eqref{A6} we have ignored the non-interacting terms such as $\hat{x}_{c}^4,\hat{x}_{m}^3$, which are negligible compared to the free Hamiltonians of the cavity and the mechanical oscillator. In addition, we have kept the interacting terms up to fourth order in $\hat{x}_c$ and $\hat{x}_m$.

The term $\alpha_m\hat{x}_m$ in Eq.~\eqref{A6} describes the qubit-induced static force, contribution of which is negligible for our system, and we neglect it.
On the other hand, the terms with coefficients $g_{Sc}$ and $g_{Sm}$ correspond to the cavity and mechanical Stark shifts, respectively.
The terms in the second line of Eq.~\eqref{A6} stand, respectively, for the radiation-pressure, cross-Kerr, cubic, and quartic couplings, each with their own coupling strength.
Finally, the terms in the third line of Eq.~\eqref{A6} can be neglected under certain conditions given in the following.

Now, we use the Bogoliubov transformation
\begin{subequations}
\begin{gather}
\hat{c}=\sinh(\theta_c)\hat{a}^\dagger+\cosh(\theta_c)\hat{a},
\\
\hat{d}=\sinh(\theta_m)\hat{b}^\dagger+\cosh(\theta_m)\hat{b},
\end{gather}
\end{subequations}
 where $\theta_{c/m}=\frac{1}{2}\ln(\omega_{c/m}^0/\sqrt{\omega_{c/m}^0(\omega_{c/m}^0+4g_{Sc/m})})$, to eliminate the Stark shift terms from Eq.~\eqref{A6}. Additionally, provided that (if $g_0\neq0$)
\begin{equation}\label{condition5}
g^0_{\rm CK},g^0_{\rm cub}, g^0_{\rm quartic}, G^0_1, G^0_2, G^0_3, G^0_4 \ll g_0,\omega_M,
\end{equation}
the terms involving different powers of the annihilation and creation operators, such as $\hat{a}^n\hat{a}^{\dagger m}$ and $\hat{b}^n\hat{b}^{\dagger m}$, rotate very fast. This implies that in the parameter regime we consider, a rotating wave approximation can be performed, and all rotating terms can be safely discarded. The surviving terms are the contributions containing equal powers of $\hat{a}$ and $\hat{a}^\dagger$, $\hat{b}$ and $\hat{b}^\dagger$.
Thus, the system Hamiltonian takes the form
\begin{gather}
\hat{H}= \omega_c\hat{a}^\dagger\hat{a}+\omega_M\hat{b^\dagger}\hat{b}+g_0\hat{a}^\dagger\hat{a}(\hat{b}^\dagger+\hat{b})
\nonumber
\\
+(g_{\rm CK}+2g_{\rm CK}'+2G_2+4G_4)\hat{a}^\dagger\hat{a}\hat{b}^\dagger\hat{b}+(g_{\rm CK}' + 2G_4)\hat{a}^\dagger\hat{a}\hat{b}^{\dagger2}\hat{b}^2
\nonumber
\\
+(G_2+2G_4)\hat{a}^{\dagger 2}\hat{a}^2\hat{b}^\dagger\hat{b}
+G_4\hat{a}^{\dagger 2}\hat{a}^2\hat{b}^{\dagger 2}\hat{b}^2.
\label{HHH}
\end{gather}
where
\begin{subequations}\label{A10}
\begin{gather}
\omega_c=\sqrt{\omega_c^0(\omega_c^0+4g_{Sc})},
\label{A10a}
\\
\omega_M=\sqrt{\omega_m^0(\omega_m^0+4g_{Sm})},
\label{A10b}
\\
g_0=2g_{rp}(\omega_c^0/\omega_c)(\omega_m^0/\omega_m)^{1/2},
\\
g_{\rm CK}=4g_{\rm CK}^0(\omega_c^0/\omega_c)(\omega_m^0/\omega_m),
\\
g_{\rm cub}=2g_{\rm cub}^0 (\omega_c^0/\omega_c)(\omega_m^0/\omega_m)^{3/2},
\\
g_{\rm CK}'= 12g^0_{\rm quartic} (\omega_c^0/\omega_c)(\omega_m^0/\omega_m)^{2},
\label{gck_def}
\\
G_2 = 12 G^0_2
(\omega_c^0/\omega_c)^{2}(\omega_m^0/\omega_m), 
\label{G2}
\\
G_4 = 36 G^0_4
(\omega_c^0/\omega_c)^{2}(\omega_m^0/\omega_m)^{2}.
\end{gather}
\end{subequations}
In the regime where the radiation pressure coupling is zero (see Sec.~\ref{section:4}), the condition \eqref{condition5} is still valid without considering $g_0$ in this relation.

The coupling strengths in Eqs.~\eqref{coefficient}
depend on the parameters $B_1,B_3,g_m,g_q$. After simplification, these parameters are given by \cite{heikkila2014}
\begin{subequations}
\begin{gather}
B_1= 2E_J,
\\
B_3= 4E_C(1-2\delta n_{g0}),
\\
g_m\approx -80\frac{E_C V_g C}{e \omega_c},
\label{gm}
\\
g_q =\frac{e^2Z_0}{4\hbar} E_J.
\label{gq}
\end{gather}
\label{E_JC}
\end{subequations}
The experimental values of $E_J,V_g,C,Z_0$ are given in Sec.~\ref{section:6}.
However, the parameters $\delta n_{\rm g0}$ and $\frac{E_J}{E_C}$ can be used as control parameters for adjusting the coupling strengths in Hamiltonian ~\eqref{HHH}.
Considering the case
\begin{equation}\label{general-condition}
G_2,G_4 << g_0, g_{\rm CK}, g_{\rm CK}',
\end{equation}
we end up with the following Hamiltonian
\begin{gather}\label{H5}
\hat{H}= \omega_c\hat{a}^\dagger\hat{a}+\omega_M\hat{b^\dagger}\hat{b}+g_0\hat{a}^\dagger\hat{a}(\hat{b}^\dagger+\hat{b})
\nonumber
\\
+\bar g_{\rm CK} \hat{a}^\dagger\hat{a}\hat{b}^\dagger\hat{b}+g_{\rm CK}'\hat{a}^\dagger\hat{a}(\hat{b}^{\dagger}\hat{b})^2,
\end{gather} 
where $\bar g_{\rm CK}=g_{\rm CK}+g_{\rm CK}'$ is the modified cross-Kerr coupling. 
Although the system parameters are restricted to satisfy $G_2 \ll g_{\rm CK}^\prime$, according to the condition of Eq.~\eqref{general-condition}, it is possible to analytically justify this condition. 
According to Eq.~\eqref{gm}, Eq.~\eqref{gq}, and also the experimental parameters (see Sec.~\ref{section:6}), we have $\abs{g_q/g_m}\approx 0.5(E_J/E_C)\ll1$. A comparison between $G_2^0$ ( Eq.~\eqref{A10G2}) and $g^0_{\rm quartic}$( Eq.~\eqref{A10g_duf}) shows that with the proper choices of the magnetic fields, $G_2^0$ can be made much smaller than $g^0_{\rm quartic}$ which in turn, according to Eqs.~\eqref{gck_def} and \eqref{G2}, leads to $G_2 \ll g_{\rm CK}^\prime$.

\section{Linearized quantum Langevin equations}\label{appendix3}
The quantum Langevin equations of motion for the Hamiltonian \eqref{H-driving} read as
\begin{subequations}
\begin{gather}\label{Langa}
\dot{\hat{a}}=-i\Delta_c \hat{a} -ig_0\hat{a}(\hat{b}^\dagger+\hat{b})-i\tilde{g}_{\rm CK}\hat{a}\hat{b}^\dagger\hat{b}
\nonumber
\\
-ig_{\rm CK}'\hat{a}\hat{b}^{\dagger2}\hat{b}^2-i\Omega -\kappa\hat{a}+\sqrt{2\kappa}\hat{a}_{\rm in},
\end{gather}
\begin{gather}\label{Langb}
\dot{\hat{b}}=-i\omega_M\hat{b}-ig_0\hat{a}^\dagger\hat{a}-i\tilde{g}_{\rm CK}\hat{a}^\dagger\hat{a}\hat{b}
\nonumber
\\
-i2g_{\rm CK}'\hat{a}^\dagger\hat{a}\hat{b}^\dagger\hat{b}^2-\gamma\hat{b}+\sqrt{2\gamma}\hat{b}_{\rm in},
\end{gather}
\end{subequations}
where $\tilde{g}_{\rm CK}=g_{\rm CK}+2g_{\rm CK}'$, and $\hat{b}_{\rm in}$ and $\hat{a}_{\rm in}$ are the corresponding input noise operators with correlation functions
\begin{subequations}
\begin{gather}\label{noise-a}
\expval{\hat{a}_{\rm in}(t)\hat{a}_{\rm in}^\dagger(t^\prime)}=\delta(t-t^\prime),
\end{gather}
\begin{equation}\label{noise-b1}
\expval{\hat{b}_{\rm in}^\dagger(t)\hat{b}_{\rm in}(t^\prime)}=\bar{n}_{\rm th}\delta(t-t^\prime),
\end{equation}
\begin{gather}\label{noise-b2}
\expval{\hat{b}_{\rm in}(t)\hat{b}_{\rm in}^\dagger(t^\prime)}=(\bar{n}_{\rm th}+1)\delta(t-t^\prime).
\end{gather}
\end{subequations}

Conversely to the Sec.~\ref{section:2}, we assume the laser driving term is intense. Then we can express the operators as the sum of classical mean values and small fluctuations, that is
$\hat{a}=\alpha+\delta\hat{a}$ and $\hat{b}=\beta+\delta\hat{b}$.
Firstly, we obtain the steady-state mean values of the cavity and resonator modes
\begin{subequations}\label{alpha-beta}
\begin{equation}\label{alpha}
(i\Delta_{\rm eff}+\kappa)\alpha-\Omega=0,
\end{equation}
\begin{equation}\label{beta}
(i\omega_{\rm eff}+\gamma)\beta+ig_0\abs{\alpha}^2=0,
\end{equation}
\end{subequations}
where
\begin{subequations}
\begin{equation}\label{tilde-deltac}
\Delta_{\rm eff}=\Delta_c+g_0(\beta+\beta^*)+\tilde{g}_{\rm CK}\abs{\beta}^2+g_{\rm CK}'\abs{\beta}^4,
\end{equation}
\begin{equation}\label{prime-omegaM}
\omega_{\rm eff}=\omega_M+\tilde{g}_{\rm CK}\abs{\alpha}^2+2g_{\rm CK}'\abs{\alpha}^2\abs{\beta}^2.
\end{equation}
\end{subequations}
We numerically solve Eqs.~\eqref{alpha-beta} to obtain the values of $\alpha$ and $\beta$ (assuming $\alpha$ is real).
The optomechanical parameters should be chosen in such a way that the conditions $\abs{\alpha},\abs{\beta}\gg1$ are satisfied to ensure that the linearization procedure is applicable. 
From numerical calculation we find $\abs{\alpha}\approx200$ and $\abs{\beta}\approx5$. Since the value of $\beta$ is not sufficiently large, we do our calculation for the case $\abs{g_{\rm CK}}\le \abs{\beta^3 g_{\rm CK}'}$, in order that the linearization approximation to be valid.
In other words, this additional condition guarantees that the generalized CK terms such as  $g_{\rm CK}'\alpha\beta^3\delta\hat{a}\delta\hat{b}$ are bigger than the CK term $g_{\rm CK}\alpha\delta\hat{a}\delta\hat{b}^\dagger\delta\hat{b}$ which is eliminated in the linearization approximation.

The dynamics of the quantum fluctuations can be described by the linearized quantum Langevin equations. Therefore in the case $\text{Im}(\beta)\ll1$, we have
\begin{subequations}\label{appendix-Langevin}
	\begin{equation}\label{appendix-Lan1}
	\delta\dot{\hat{a}}=-(i\Delta_{\rm eff}+\kappa)\delta\hat{a}+i\frac{g_{\rm eff}}{2}(\delta\hat{b}^\dagger+\delta\hat{b})+\sqrt{2\kappa}\hat{a}_{\rm in},
	\end{equation}
	\begin{equation}\label{appendix-Lan2}
	\delta\dot{\hat{b}}=-(i\omega_{\rm eff}+\gamma)\delta\hat{b}+i\frac{g_{\rm eff}}{2}(\delta\hat{a}^\dagger+\delta\hat{a})+\sqrt{2\gamma}\hat{b}_{\rm in},
	\end{equation}
\end{subequations}
where
\begin{equation}\label{G}
g_{\rm eff}=-2g\abs{\alpha}, \qquad g=g_0+\tilde{g}_{\rm CK}\beta + 2g_{\rm CK}'\beta^3.
\end{equation}
\section{Routh-Hurwitz criterion}\label{appendix4}
The Routh-Hurwitz criterion for the stability of the system gives the following four independent conditions
\begin{subequations}\label{condition}
	\begin{equation}\label{condition1}
	2\gamma+2\kappa>0,
	\end{equation}
	\begin{equation}\label{condition2}
	\kappa\Delta_{\rm eff}^2+\gamma\omega_{\rm eff}^2+\kappa^3+\gamma^3+4\kappa\gamma^2+\kappa^2\gamma>0,
	\end{equation}
	\begin{eqnarray}\label{condition3}
	4\kappa\gamma[\Delta_{\rm eff}^4&&+2\Delta_{\rm eff}^2(\gamma^2+\kappa^2-\omega_{\rm eff}^2)+4\kappa\gamma(\Delta_{\rm eff}^2+\kappa\gamma+\gamma^2+\kappa^2+\omega_{\rm eff}^2)
	\nonumber
	\\
	&&+(\gamma^2+\kappa^2+\omega_{\rm eff}^2)^2
	]+4g_{\rm eff}^2\Delta_{\rm eff}\omega_{\rm eff}(\gamma+\kappa)^2>0,
	\end{eqnarray}
	\begin{equation}\label{condition4}
	(\gamma^2+\omega_{\rm eff}^2)(\kappa^2+\Delta_{\rm eff}^2)>\omega_{\rm eff}g_{\rm eff}^2\Delta_{\rm eff}.
	\end{equation}
\end{subequations}
While the first two conditions are trivial, the two other ones are nontrivial. If we restrict the Hamiltonian~\eqref{H-linear-effective}  to the regime $\Delta_{\rm eff}>0$, then the third condition will always be satisfied. Note that the last condition depends on the system parameters, and hence should be satisfied.


\end{document}